\documentclass[%
 reprint,
 amsmath,amssymb,
 aps,
 prl,
]{revtex4-2}

\usepackage{graphicx}
\usepackage{dcolumn}
\usepackage{bm}
\usepackage[dvipsnames]{xcolor}

\newcommand{\ion}[1]{#1_{\text{i}}}
\newcommand{\proton}[1]{#1_{\text{p}}}

\newcommand{\protongk}{\rho_{\text{p,gk}}}
\newcommand{\deltau}{\delta u}
\newcommand{\deltaui}{\deltau_{\ion{\rho}}}
\newcommand{\deltaup}{\deltau_{\proton{\rho}}}
\newcommand{\deltaulambda}{\deltau_\lambda}
\newcommand{\vthi}{v_{\text{th,i}}}
\newcommand{\vthp}{v_{\text{th,p}}}
\newcommand{\vA}{v_{\text{A}}}
\newcommand{\vph}{v_{\text{ph}}}

\newcommand{\Espec}{\mathcal{E}_{\bm{E}}(k_\perp)}
\newcommand{\zpspec}{\mathcal{E}^+(k_\perp)}
\newcommand{\ExB}{\bm{E}\times\bm{B}}
\newcommand{\nlfreq}{k_\perp \tilde{z}^+_{k_\perp}}

\newcommand{\explicitxi}{\deltaui/\vthi}
\newcommand{\explicitxiproton}{\deltaup/\vthp}
\newcommand{\kprptilde}{\tilde{k}_\perp}
\newcommand{\kprltilde}{\tilde{k}_\|}

\newcommand{\flrbal}{$\text{F-Bal-}\proton{\beta}0.05$}
\newcommand{\rmhd}{$\text{R-}\proton{\beta}0.05$}
\newcommand{\rmhdHe}{$\text{R-}\ion{\beta}0.05\text{He}^{2\text{+}}$}
\newcommand{\rmhdO}{$\text{R-}\ion{\beta}0.05\text{O}^{5\text{+}}$}
\newcommand{\flrimbalthree}{$\text{F-3}\proton{\beta}0.05$}
\newcommand{\flrimbalsix}{$\text{F-6}\proton{\beta}0.05$}
\newcommand{\flrimbalsixlowbeta}{$\text{F-6}\proton{\beta}0.01$}
\newcommand{\flrimbalsixhighbeta}{$\text{F-6}\proton{\beta}0.1$}
\newcommand{\flrimbalsixHe}{$\text{F-6}\ion{\beta}0.05\text{He}^{2+}$}
\newcommand{\flrimbalsixO}{$\text{F-6}\ion{\beta}0.05\text{O}^{5+}$}
\newcommand{\flrimbalten}{$\text{F-10}\proton{\beta}0.05$}

\begin{document}

\preprint{APS/123-QED}

\title{A Unified Phenomenology of Ion Heating in Low-$\beta$ Plasmas: Test-Particle Simulations}

\author{Zade Johnston}
 \email{johza721@student.otago.ac.nz}
\author{Jonathan Squire}%
\author{Romain Meyrand}
\affiliation{%
 Physics Department, University of Otago, Dunedin 9010, New Zealand
}%
\date{\today}

\begin{abstract}
We argue that stochastic and resonant ion heating, often viewed as distinct processes in low-$\beta$ collisionless plasmas, are the far limits of a continuum controlled by nonlinear broadening of turbulent fluctuations, and thus by the normalized cross helicity.
We propose a simple empirical formula that captures both regimes, generalizing that commonly used to describe stochastic heating.
Simulations of test ions interacting with turbulence confirm our scalings across a wide range of different ion and turbulence properties, including with a steep ion-kinetic transition range as seen in the solar wind.
Our results provide a unified framework for understanding ion heating processes across diverse astrophysical environments from black-hole accretion disks to the solar corona, also providing a compact and versatile subgrid model for larger-scale simulations.
\end{abstract}

\maketitle

Turbulent ion heating is a key process in collisionless astrophysical plasmas such as the solar corona, accretion disks, the hot interstellar medium, and the intracluster medium \cite[e.g.,][]{Quataert1999-rq,Ferriere2001,Kunz2022,Howes2024-fn}.
How turbulent energy is apportioned among species and channels sets the plasma's temperature, structure, radiative efficiency, and large-scale dynamical evolution.
In the well-studied solar wind and corona, ions are observed to be continuously heated, with more energization perpendicular to the local magnetic field than parallel \cite{Marsch1982-vu,Marsch2004-fw,Hellinger2006-ff}, a process that plays a crucial role in driving solar wind acceleration and sustaining the high temperatures of the corona.
Heavy ions are heated more than protons \cite{Kohl1998-cq,Esser1999-qd,Antonucci2000-lz} and protons more than electrons \cite{Cranmer2009-vf}.
Many mechanisms have been proposed to explain these observations, including stochastic heating by uncorrelated turbulent fluctuations \cite{Chandran2010-ow,Chandran2013-ng}, wave-particle interactions leading to resonant heating \cite{Hollweg2002-dw,Isenberg2011-wt,Isenberg2019-oy}, and others \cite{Chandran2005-mb,Schekochihin2009-qo,TenBarge2013-ux}.
However, theories of ion heating processes remain incomplete.
As well as hindering our understanding of the heliosphere, this crucial deficiency represents an important limitation for modeling the dynamics and energetics of a wide array of other astrophysical processes.

Extending previous studies \cite{Lehe2009-up,Xia2013-ob,Teaca2014-iq,Weidl2015-sd,Arzamasskiy2019-qv,Cerri2021-xo,Pugliese2023-px}, we present a unified description of ion heating in turbulence.
We argue that the normalized cross helicity (or ``imbalance")
\begin{equation}
    \sigma_{\rm c}= \frac{\langle (\bm{z}^+)^2 - (\bm{z}^-)^2 \rangle}{\langle (\bm{z}^+)^2 + (\bm{z}^-)^2 \rangle},
    \label{eq:imbalance}
\end{equation}
the relative proportion of fluctuations propagating outwards ($\bm{z}^+$) versus inwards ($\bm{z}^-$) in the turbulence, is of key importance.
Specifically, because imbalance controls the wavenumber-frequency spectrum of the turbulence, increasing it will induce a change from stochastic heating (SH) to resonant heating (RH) as fluctuations transition from broadband to being more sharply peaked in frequency.
We further propose that the standard empirical formula for the heating rate in SH, regularly applied for empirical modeling and interpreting observations \cite{Bourouaine2013-gh,Klein2016-qo,Vech2017-jq,Martinovic2019-xf,Martinovic2020-do}, can be extended to also cover imbalanced turbulence with a transition-range break in the spectrum, with similar heating efficiencies in both cases.
We test this proposal with high-resolution simulations of test particles interacting with low-$\beta$ turbulence.
The scaling accurately captures measured proton and minor-ion heating rates in turbulence across a range of conditions (balanced, imbalanced, with and without a ``helicity barrier").
These results help clarify the physics of collisionless turbulent ion heating, with clear application to both in-situ observational analyses of the near-Sun environment and broader theoretical models of astrophysical plasmas \cite{Howes2024-fn}.

We first define the following symbols, with species index $s$ either ions, protons, or electrons ($s=\rm i, \ p, \ or \ e$):
$\bm{E}$ and $\bm{B}$ are the electric and magnetic fields; $k_\|$ and $k_\perp$ are the components of the wavevector $\bm{k}$ parallel and perpendicular to $\bm{B}$; $q_s = Ze$ and $m_s = A m_\text{p}$ are the ion charge and mass in units of proton quantities; $c$ and $\vA$ are the speed of light and the Alfv\'en speed; $\rho_s = v_{\text{th},s}/\Omega_s$ is the thermal gyroradius with $v_{\text{th},s} = \sqrt{2T_s/ m_s}$ and $\Omega_s=|q_s| B/m_s c$ the ion thermal speed and gyrofrequency; $d_s = \rho_s / \sqrt\beta_s$ is the ion inertial length where $\beta_s = v^2_{\text{th},s} / \vA^2$ is the ion beta; and $\deltaulambda$ is the rms $\ExB$ velocity at scales $k_\perp\lambda\sim 1$.

\textit{Ion Heating in Turbulence.}---Two widely studied ion-heating mechanisms are stochastic and resonant heating.
SH involves ions heated by uncorrelated kicks from fluctuations at scales $k_\perp\ion{\rho}\sim 1$, leading to a diffusion perpendicular to $\bm{B}$ in velocity space and generally heating the plasma \cite{McChesney1987-dp,Johnson2001-ui,Chandran2010-ow}.
In contrast to this non-resonant process, RH arises from strong ion-wave interactions when the waves' Doppler-shifted frequencies are resonant with $\ion{\Omega}$ \cite{Kennel1966-rx,Stix1992-bo,Schlickeiser1993-if}.
This causes ions to diffuse in velocity space along contours of constant energy in the frame moving at the wave's phase velocity, heating the plasma.
Both theories have seen support from simulations \cite{Xia2013-ob,Hoppock2018-fv,Arzamasskiy2019-qv,Isenberg2019-oy,Cerri2021-xo,Zhang2025-yd} and observations of the near-Sun environment \cite{Bourouaine2013-gh,Klein2016-qo,Vech2017-jq,Martinovic2019-xf,Martinovic2020-do,Bowen2022-qt,Bowen2024-si}.

While traditionally considered as separate mechanisms, we argue that SH and RH should instead be considered as two limits of a continuum controlled by the nonlinear broadening of the frequency spectrum of turbulent fluctuations.
A measure of this spectrum is the space-time Fourier transform $\mathcal{E}^{\rm tot}(\bm{k},\omega)=\mathcal{E}^+(\bm{k}, \omega) + \mathcal{E}^-(\bm{k}, \omega)$, where $\mathcal{E}^{\pm}(\bm{k}, \omega) = \frac{1}{2}|\bm{z}^{\pm}(\bm{k}, \omega)|^2$, which gives the power in frequencies at a given $\bm{k}$.
Unbroadened fluctuations---waves with a single frequency $\omega(\bm{k})$ at a particular $\bm{k}$ such that $\mathcal{E}^{\rm tot}(\bm{k},\omega)$ is sharply peaked (e.g., $\propto\delta\left[\omega - \omega(\bm{k})\right]$)---cause RH-like behavior; in contrast, SH-like behavior is caused by fluctuations that are broadened to a level comparable to $\omega$ itself, which thus have significant power at low-frequencies \cite{Schekochihin2022-nn}.
This reveals a deep connection to the turbulence imbalance: in imbalanced turbulence $\bm{z}^+$-fluctuations have linear frequencies that exceed their nonlinear interaction frequency (the RH-limit), whereas in balanced turbulence the two are comparable \cite{Goldreich1995-fv} (the SH-limit).
A smooth transition between these limits occurs as the imbalance is adjusted.

In testing this idea with theory and simulations below, a complication is the ``helicity barrier" (HB) \cite{Meyrand2021-ix,Squire2022-dm,Squire2023-jn}.
This occurs in imbalanced turbulence in collisionless plasmas, halting the cascade and creating a steep drop in power and $\ion{\rho}$ (the ``transition range").
While the HB does not fundamentally change our arguments, it indirectly affects the frequencies of fluctuations that most strongly heat ions, and so must be considered.
The HB is described further in the End Matter.

\textit{Unified Ion Heating Formula}---We now argue that SH and RH rates scale similarly with turbulent amplitude. 
SH requires $\deltaui$, the turbulent amplitude at $\ion{\rho}$-scales, to be comparable to $\vthi$ (generally  $\explicitxi\gtrsim 0.1$ \cite{Chandran2010-ow,Xia2013-ob}).
To understand the amplitude at which RH will be activated in imbalanced turbulence, we use the principle of ``propagation critical balance" (where parallel scales are determined by the wandering of magnetic field lines by perpendicular perturbations \cite{Lithwick2007-ao,Beresnyak2008-aw,Schekochihin2022-nn}), giving $\omega\sim k_\|\vA \sim \deltaulambda/\lambda$ throughout the inertial range.
Thus in order to reach $\omega\gtrsim\ion{\Omega}$, we require amplitudes such that $\deltaulambda/\lambda \gtrsim \ion{\Omega} = \vthi/\ion{\rho}$ and thus $\deltaulambda/\vthi\gtrsim \lambda/\ion{\rho}$.
Taking $\lambda\sim\ion{\rho}$ shows that heating will occur for turbulent amplitudes similar to those needed for SH, although in principle we can have heating at scales $\lambda > \ion{\rho}$ (we ignore interactions with $k_\perp\ion{\rho}>1$ fluctuations because they are suppressed by particles sampling many small-scale eddies over an orbit \cite{Chandran2000-pq}; however, see \cite{Arzamasskiy2019-qv,Isenberg2019-oy}).
Our argument shows that RH is activated at similar turbulent amplitudes to SH.
Similarly, for critically balanced turbulence, the amplitudes $\deltaui\sim\vthi$ needed to cause SH imply turbulent frequencies approaching $\ion{\Omega}$ at the edge of the critical balance cone, with a flat spectrum at lower frequencies \cite{Schekochihin2022-nn}; this occurs despite SH traditionally being discussed in the context of low-frequency turbulence ($\omega\ll \ion{\Omega}$).

For SH, Ref.~\cite{Chandran2010-ow} argue that the ion-heating rate per unit mass is
\begin{equation}
    Q_\perp = \ion{\Omega} \vthi^2 c_1 \ion{\xi}^3 F(\ion{\xi}).
    \label{eq:Qprp}
\end{equation}
Here $\ion{\xi} = \deltaui/\vthi$, $F(\ion{\xi})$ is a ``suppression factor" that Ref.~\cite{Chandran2010-ow} propose as $F(\ion{\xi})=e^{-c_2/\ion{\xi}}$ to account for the magnetic moment being exponentially conserved, and $c_1$ and $c_2$ are empirically determined constants.
In the End Matter we show, through application of quasi-linear theory \cite{Kennel1966-rx}, that a heating rate of the same form as Eq.~\eqref{eq:Qprp} emerges for a simple model of critically-balanced turbulence in the imbalanced regime.

We thus propose to extend  Eq.~\eqref{eq:Qprp} to cover all turbulence regimes---balanced (with SH) and imbalanced (with RH).
For simplicity, we use $F(\ion{\xi})=e^{-c_2/\ion{\xi}}$ in all cases, noting that although this choice is not formally justified, it works well empirically in matching our simulations (however, we do not rule out a weaker dependence of $F(\ion{\xi})$ on imbalance or other parameters, which is expected theoretically; see End Matter).

For turbulence without a HB (e.g., balanced turbulence), the energy cascade is able to dissipate at perpendicular scales smaller than $\ion{\rho}$, forming a smooth spectrum with continuously increasing $\deltaulambda/\lambda$.
Based on the argument above, we take $\ion{\xi}$ to be the standard SH measure $\explicitxi$.
In contrast, the presence of a HB causes a drop in the spectrum at scales $\lambda > \ion{\rho}$, suggesting the frequency of fluctuations reach a maximum at or near this scale.
Because the heating in this regime does not depend specifically on $\ion{\rho}$-scale fluctuations, but rather the power in fluctuations at scales $\lambda\gtrsim\ion{\rho}$ with frequencies approaching $\ion{\Omega}$, we set $\ion{\xi}\equiv (\ion{\rho}/\lambda)^{1/3} \deltaulambda/\vthi$, motivated heuristically as being the parameter combination that ensures Eq.~\eqref{eq:Qprp} has no explicit dependence on $\ion{\rho}$ and justified more rigorously by the analytic quasi-linear calculation in the End Matter.
We will take $\lambda$ to be the perpendicular scale at which fluctuations exhibit their highest frequency.
This formula---Eq.~\eqref{eq:Qprp} with a modified $\ion{\xi}$---also takes into account that minor ions interact with fluctuations at scales larger than the transition range, due to their larger gyroradii.

To verify this proposal, we now present numerical simulations of test protons and minor ions interacting with both balanced and imbalanced turbulence with or without a HB.

\textit{Numerical Setup.}---We use the ``finite Larmor radius MHD" (FLR-MHD) model \cite{Passot2018-yt, Schekochihin2019-al,Meyrand2021-ix}, which can be formally derived from gyrokinetics in the limit $\beta_\perp\sim\beta_\|\sim\beta_{\rm e} \ll 1$ and $k_\perp d_{\rm e}\ll 1$.
At $k_\perp\protongk\ll 1$  (we assume a majority hydrogen plasma, with $\protongk$ the $\ion{\rho}$ in the gyrokinetic model), FLR-MHD reduces to the well-known RMHD model \cite{Strauss1976-vu}; for $k_\perp\protongk\gg 1$ it reduces to ``electron RMHD" \cite{Schekochihin2009-qo,Boldyrev2013-gq,Cerri2021-xo}.
These equations are advanced in time using a pseudospectral method \cite{Teaca2009-go,Meyrand2021-ix} (see Supplemental Material). The simulation domain is a three-dimensional periodic grid of size $L_\perp = L_z=2\pi$ and a resolution of $N_\perp^2\times N_z$, with background magnetic field $\bm{B}_0=B_0\hat{\bm{z}}$.
To test our model across a wide range of regimes, we simulate three cases: balanced and imbalanced FLR-MHD with $\protongk/L_\perp = 0.02$, and imbalanced RMHD.
These simulations have a resolution of $N_\perp^2\times N_z = 1024^2\times 1024$ for the balanced FLR-MHD case and $1024^2\times 512$ for the other two cases, refined from an initial resolution of $256^2\times 512$ to obtain spectra that extend to scales smaller than $\protongk$ (see Supplemental Material).
In the imbalanced FLR-MHD simulations we compare three different stages of the HB's growth, at times $t\vA/L_z\approx 3, 6,$ and 10.

\begin{table}[b]
\caption{\label{tab:sim-names} Simulation parameters studied in this Letter. F and R correspond to the FLR-MHD and RMHD models. Coefficients $c_1,c_2$ and $\hat{c}_1,\hat{c}_2$ are fits to Eq.~\eqref{eq:Qprp} using $\explicitxi$ and our proposed $\ion{\xi}$; the best fit of all data using $\ion{\xi}$ is $\hat{c}_1,\hat{c}_2=2.09,0.13$.}
\begin{ruledtabular}
\begin{tabular}{lcccc}
\textrm{Name}&
$\ion{\beta}$&
$A/Z$&
$c_1, c_2$&
$\hat{c}_1, \hat{c}_2$\\
\colrule
\flrbal              & 0.05 & 1    & 3.46, 0.21  & 3.91, 0.20\\
\rmhd                & 0.05 & 1    & 1.66, 0.19  & 2.08, 0.18\\
\rmhdHe              & 0.05 & 4/2  & 2.50, 0.30  & 2.62, 0.29\\
\rmhdO               & 0.05 & 16/5 & 2.42, 0.18  & 2.55, 0.17\\
\flrimbalthree       & 0.05 & 1    & 19.30, 0.08  & 4.85, 0.13\\
\flrimbalsix         & 0.05 & 1    & 30.49, 0.07 & 3.06, 0.14\\
\flrimbalsixlowbeta  & 0.01 & 1    & 40.33, 0.08 & 4.03, 0.16\\
\flrimbalsixhighbeta & 0.1  & 1    & 19.26, 0.06 & 1.94, 0.12\\
\flrimbalsixHe       & 0.05 & 4/2  & 5.08, 0.16  & 3.81, 0.18\\
\flrimbalsixO        & 0.05 & 16/5 & 2.45, 0.14  & 2.83, 0.13\\
\flrimbalten         & 0.05 & 1    & 69.9, 0.06  & 3.56, 0.15\\
\end{tabular}
\end{ruledtabular}
\end{table}

Once the refined turbulence simulations reach a quasi-steady-state, we introduce test ions (see Supplemental Material).
For a given collection of ions, we choose the initial $\ion{\rho}/L_\perp$, $\ion{\beta}$, and $\explicitxi$ \cite{Xia2013-ob}.
In FLR-MHD, the choice of $\ion{\rho}/L_\perp$ is fixed to match $\protongk/L_\perp=0.02$; for comparison, we also choose this value in the RMHD simulations.
This scale determines $\deltaui$ via $\deltaui^2 = 2(c/B_0)^2\int_{k^-}^{k^+} \text{d}k_\perp \ \Espec$, 
where $\Espec$ is the perpendicular electric-field spectrum (equivalent to the $\ExB$ 
velocity spectrum in FLR-MHD) and using $k^\pm \equiv e^{\pm1/2}/\ion{\rho}$ \cite{Chandran2010-ow,Xia2013-ob}.
This value of $\deltaui$ with the choice of $\explicitxi$ determines $\vthi$.
Due to the invariance of the quantity $\vA\nabla_\|$ in both RMHD and FLR-MHD, we are free to choose $\vA$ based on the values of $\vthi$ and $\ion{\beta}$ so long as length-scales along $\bm{B}_0$ are scaled by the same factor \cite{Schekochihin2022-nn}; this rescaling is done within the ion integrator.
The ions are then uniformly distributed throughout the domain and given an isotropic Maxwellian velocity distribution.
For a given choice of $\ion{\rho}/L_\perp$ and $\ion{\beta}$, we initialize separate cohorts of $N=10^6$ particles with $0.01 \lesssim \explicitxi \lesssim 0.3$, a range relevant to the solar wind and corona.

We introduce minor ions with mass $\ion{m}=A\proton{m}$ and charge $\ion{q}=Ze$ by taking $\vthi$ to be equal to that of protons within the same simulation, so that they have similar initial $\explicitxi$.
This choice, along with $\ion{\Omega}=(Z/A)\proton{\Omega}$, determines the required gyroradius $\ion{\rho} = (A/Z)\proton{\rho}$.
The minor ions are then initialized identically to protons.
Table \ref{tab:sim-names} summarizes the turbulence and particle parameters studied in this Letter.

Because FLR-MHD is derived from gyrokinetics, it can only simulate low-frequency dynamics and excludes the ion inertial length $\ion{d}$.
Although this still allows for RH of protons by Alfv\'enic fluctuations, it does not capture their dispersion as they transition into ion-cyclotron waves at $k_\| \ion{d}\sim 1$, as in previous MHD-based studies \cite{Xia2013-ob,Teaca2014-iq,Martinovic2020-do,Pugliese2023-px} (a hybrid-kinetic model is needed to capture this correctly \cite{Squire2022-dm}).
The true resonance with ion-cyclotron waves will be at higher $k_\|$, but due to critical balance these modes will have less power than in FLR-MHD.
Minor ions, with their lower gyrofrequencies, are less affected by this approximation.

\begin{figure}
\includegraphics[scale=0.6]{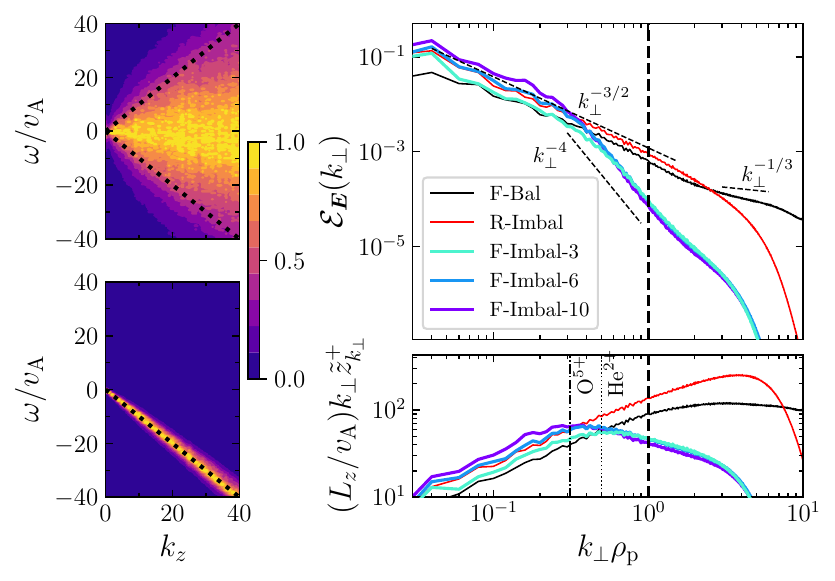}
\caption{\label{fig:spectra} Left: Space-time Fourier transforms $\mathcal{E}^{\rm tot}(k_z, \omega)$ from balanced (top) and imbalanced (bottom) FLR-MHD turbulence, normalized to the maximum value at each $k_z$; the dotted lines are the Alfv\'en wave dispersion relation $\omega=\pm k_z\vA$. Right: Comparison of electric-field spectra (top) and the nonlinear frequency} $\nlfreq$ (bottom) used to estimate $\lambda\lesssim \proton{\rho}$ (the scale at which $\nlfreq$ reaches a maximum) from the turbulence simulations in this Letter.
\end{figure}

\begin{figure*}
\includegraphics[width=\textwidth]{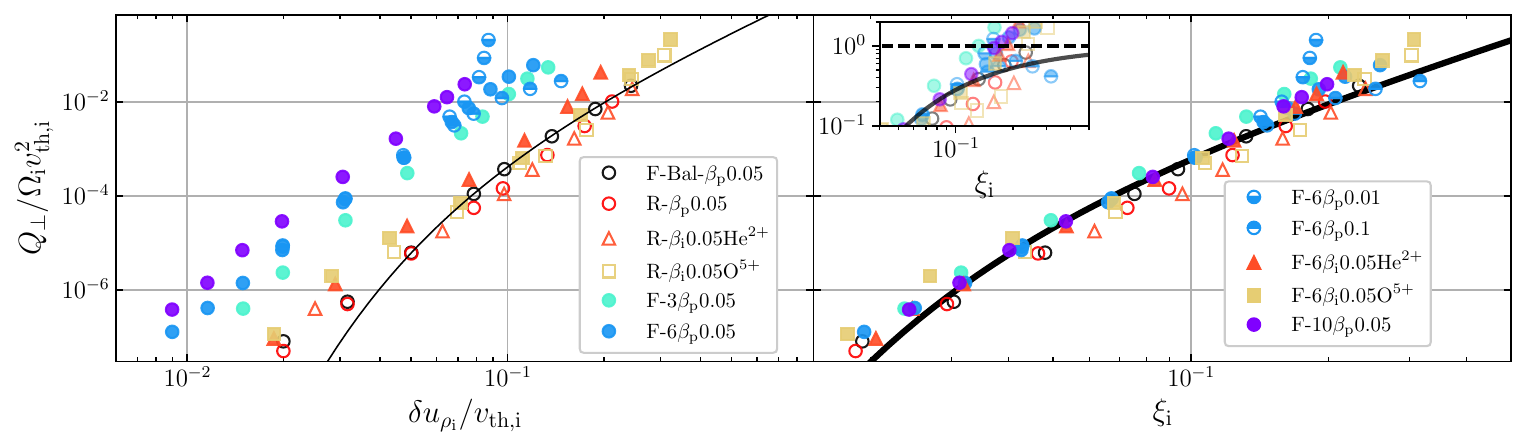}
\caption{\label{fig:SH_heatingrate} Comparison of the measured $Q_\perp$ against the commonly used SH parameter $\explicitxi$ (left) and our proposed $\ion{\xi}$ (right). Filled/hollow markers represent simulations with/without a HB. The inset shows $Q_\perp / (\hat{c}_1\ion{\xi}^3)$, with $\hat{c}_1=2.09$ the best fit to Eq.~\eqref{eq:Qprp} for all data, highlighting the exponential suppression for small $\ion{\xi}$ (deviation from $Q_\perp\propto \ion{\xi}^3$, dashed). The solid black lines show the best fit of Eq.~\eqref{eq:Qprp} to \flrbal~(left) and all data (right and inset).}
\end{figure*}

\textit{Results.}---To highlight how imbalance affects the frequency spectrum of fluctuations, which was argued above to be important in controlling ion heating,
Fig.~\ref{fig:spectra} shows the $k_\perp$-averaged space-time Fourier transform $\mathcal{E}^{\rm tot}(k_z,\omega)$ of FLR-MHD turbulence.
We see that imbalanced turbulence is dominated by fluctuations peaked around the Alfv\'en frequency $\omega \approx -k_z\vA$.
In contrast, $\mathcal{E}^{\rm tot}(k_z,\omega)$ has more power in fluctuations with $|\omega|/(k_z\vA)\ll 1$ in balanced turbulence due to the nonlinear broadening of fluctuations, a signature that the nonlinear and linear times are comparable \cite{Schekochihin2022-nn}.
As argued above, this change in the frequency spectrum of the turbulence causes the ion heating to transition from a SH-like mechanism in the balanced case to a RH-like mechanism in the imbalanced case.

The electric-field spectra, $\Espec$, and the nonlinear frequency of fluctuations, $\nlfreq$, for different regimes are also compared in Fig.~\ref{fig:spectra},
where we compute $\nlfreq$ from its spectrum: $\nlfreq \approx k_\perp\sqrt{k_\perp\zpspec}$ \cite{Bott2021-vp}.
The balanced FLR-MHD simulation captures the transition from an Alfv\'en-wave to kinetic-Alfv\'en-wave cascade \cite{Schekochihin2009-qo}, with $\Espec$ approaching a $k_\perp^{-1/3}$ scaling at $k_\perp\proton{\rho}\gtrsim 1$.
The presence of a HB is clearly seen in the imbalanced FLR-MHD simulations via the steep drop in spectra at $k_\perp\proton{\rho}\lesssim 1$, as observed in the solar wind \cite{Kiyani2015-yk,Bowen2020-wa} and previous simulations \cite{Meyrand2021-ix,Squire2022-dm,Squire2023-jn}.
The steep spectral slope of the HB causes $\nlfreq$ to reach a maximum at $k_\perp\proton{\rho}\lesssim 1$.
This maximum, and the corresponding spectral break in $\Espec$, shifts to larger scales as the energy grows in time.

Figure \ref{fig:SH_heatingrate} compares the dependence of the perpendicular heating rate of the test particles, $Q_\perp$, on $\explicitxi$, the commonly used SH measure, to our proposed definition $\ion{\xi}=(\ion{\rho}/\lambda)^{1/3} \deltaulambda/\vthi$, with $\deltaulambda$ and $\vthi$ taken at $t_0$; see End Matter for details on the numerical calculation of $Q_\perp$.
We calculate $\lambda=1/k^{\rm max}_\perp$ by using the scale $k^{\rm max}_\perp < 1/\ion{\rho}$ for which $\nlfreq$ reaches its maximum in Fig.~\ref{fig:spectra}.
The standard SH measure $\explicitxi$ is seen to be inadequate for describing $Q_\perp$ for cases with a HB, which show greater-than-expected heating from the power in fluctuations at $\ion{\rho}$-scales indicating the true $\ion{\xi}$ is being underestimated.
Minor ions show similar $Q_\perp$ in the imbalanced FLR-MHD and RMHD simulations, as their gyroradii lie at scales larger than the transition-range drop.
When plotted against our proposed $\ion{\xi}$, the data are well described by Eq.~\eqref{eq:Qprp} with best fit parameters $\hat{c}_1 = 2.09,\hat{c}_2 = 0.13$ (fits of individual simulations to Eq.~\eqref{eq:Qprp} are listed in Table \ref{tab:sim-names}); our value of $\hat{c}_2$ is smaller than previous test particle simulations \cite{Chandran2010-ow,Xia2013-ob} but similar to recent hybrid-kinetic simulations \cite{Cerri2021-xo}.
That Eq.~\eqref{eq:Qprp} holds despite the difference in simulation conditions, from turbulence that is balanced, imbalanced, with or without a HB, and for protons and minor ions, helps verify our unified model of ion heating.
We note the worst agreement is seen for imbalanced RMHD, which is unphysical (it neglects the effect of $\ion{\rho}$ on the turbulence), and that the fit at larger $\ion{\xi}$ is improved slightly if one measures $\vthi$ from the initial distribution, rather than at $t_0$ (as in Refs.~\cite{Chandran2010-ow,Xia2013-ob,Hoppock2018-fv}).

\begin{figure}
\includegraphics[scale=0.75]{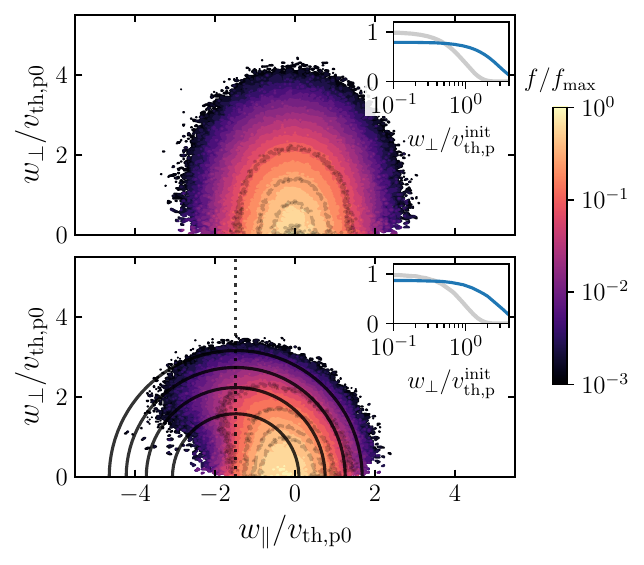}
\caption{\label{fig:SH_veldists} Comparison of the VDFs $f(w_\perp, w_\|)$ of $\proton{\hat{\xi}}\approx0.3$ protons from \flrbal~(top) and $\proton{\hat{\xi}}\approx0.7$ protons from \flrimbalsixhighbeta~(bottom), with velocities scaled by $\vthp$ at $t_0$. Dashed contours represent the distribution at $t_0$, and solid contours represent constant-energy contours in the wave frame, centered on the phase speed $\vph=-\vA$ (dotted line). Insets show the 1D VDF $f(w_\perp)$ (blue), compared to the initial Maxwellian distribution (gray).}
\end{figure}

Despite the success of the unified model in reproducing $Q_\perp$, there remain important differences between balanced- and imbalanced-turbulent heating.
Figure \ref{fig:SH_veldists} compares the proton velocity distribution functions (VDFs) $f(w_\perp, w_\|)$ in these regimes.
In balanced turbulence, the VDF undergoes diffusion in perpendicular velocity, as seen in the 1D VDF $f(w_\perp) = \int \text{d}w_\| f(w_\perp, w_\|)$ which has a characteristic flattopped shape \cite{Klein2016-qo,Cerri2021-xo}.
In imbalanced turbulence the VDF diffuses along constant-energy contours within the frame of the waves, characteristic of RH \cite{Kennel1966-rx} (for Alfv\'enic fluctuations, these contours are semicircles centered on the phase speed $\vph = \omega/k_\| = -\vA$).
The 1D VDF exhibits a flattop similar to the balanced case (dropping off slightly earlier upon close inspection), presumably because the contours are nearly vertical at small $w_\perp$.
This picture clarifies the $\proton{\beta}$-dependent decrease in $Q_\perp$ at large $\ion{\xi}$ in Fig.~\ref{fig:SH_heatingrate} (representing the largest deviation from the model): the constant-energy contours become increasingly curved as $\proton{\beta}$ increases, causing ions to gain parallel energy at the expense of perpendicular energy (see Supplemental Material).

\textit{Conclusion.}---We provide a unified picture for describing ion heating in collisionless plasma turbulence across a wide range of regimes.
The heating mechanism changes character depending on the frequency spectrum of turbulent fluctuations, transitioning from SH to RH as the turbulence imbalance increases and reduces the relative nonlinear broadening of fluctuations.
We propose to extend a commonly used empirical model of heating in balanced turbulence to also capture imbalanced turbulence with a transition-range break caused by the HB, which causes a steep drop in turbulent energy at $k_\perp\proton{\rho} \sim 1$.
Using high-resolution simulations of test particles interacting with turbulence, we show that our proposal works well in describing measured heating rates for both protons and minor ions in turbulence that is balanced, imbalanced, with or without a HB, even though 
the form of the ion VDF that results from the heating has a strong dependence on imbalance.

The FLR-MHD model, on which our results are based, is simplified, omitting the transition to ion-cyclotron waves when $\omega\sim\ion{\Omega}$.
Additionally, we only measure the instantaneous $Q_\perp$ from an initial Maxwellian rather than tracking large changes of the VDF (as occurs for strong heating); different choices of initial velocity distribution would also affect the measured heating rate, as SH and RH presumably saturate in different ways.
Given their modest scale separation, our results may also be affected by intermittency effects; however, this dependence is likely weak \cite{Xia2013-ob}.
The issues of missing ion-cyclotron wave physics and long-term heating are addressed in a recent hybrid-kinetic study \cite{Zhang2025-yd}, which shows similar features in their VDFs.
They also observe similar minor ion heating rates in balanced and imbalanced turbulence with similar amplitudes during the initial stages of heating, consistent with our results.

The results of this Letter, especially the link between SH, RH, and turbulence imbalance, help to disentangle signatures of ion heating in astrophysical environments, providing a base for more complex theories.
For instance, it is noteworthy that ion heating rates are similar in balanced and imbalanced turbulence at the same amplitude given that their turbulent cascade rates differ; this has interesting implications for the level at which turbulence saturates given some driving force.  Developing these ideas, especially with the effects of the HB included, is needed for understanding the global dynamics and observational signatures of diverse astrophysical processes.

\begin{acknowledgments}
The authors would like to thank T.~Adkins, B.~D.~G.~Chandran, M.~W.~Kunz, N.~F.~Loureiro, M.~Zhang, and M.~Zhou for helpful discussions over the course of this work.
Research is supported by the University of Otago, through a University of Otago Doctoral Scholarship (ZJ), and the Royal Society Te Ap\=arangi, through Marsden-Fund grant MFP-UOO2221 (JS) and MFP-U0020 (RM), as well as through the Rutherford Discovery Fellowship RDF-U001004 (JS).
High-performance computing resources were provided by the New Zealand eScience Infrastructure (NeSI) under project grant uoo02637. 
\end{acknowledgments}

\bibliography{main}

\begin{thebibliography}{70}%
\makeatletter
\providecommand \@ifxundefined [1]{%
 \@ifx{#1\undefined}
}%
\providecommand \@ifnum [1]{%
 \ifnum #1\expandafter \@firstoftwo
 \else \expandafter \@secondoftwo
 \fi
}%
\providecommand \@ifx [1]{%
 \ifx #1\expandafter \@firstoftwo
 \else \expandafter \@secondoftwo
 \fi
}%
\providecommand \natexlab [1]{#1}%
\providecommand \enquote  [1]{``#1''}%
\providecommand \bibnamefont  [1]{#1}%
\providecommand \bibfnamefont [1]{#1}%
\providecommand \citenamefont [1]{#1}%
\providecommand \href@noop [0]{\@secondoftwo}%
\providecommand \href [0]{\begingroup \@sanitize@url \@href}%
\providecommand \@href[1]{\@@startlink{#1}\@@href}%
\providecommand \@@href[1]{\endgroup#1\@@endlink}%
\providecommand \@sanitize@url [0]{\catcode `\\12\catcode `\$12\catcode `\&12\catcode `\#12\catcode `\^12\catcode `\_12\catcode `\%12\relax}%
\providecommand \@@startlink[1]{}%
\providecommand \@@endlink[0]{}%
\providecommand \url  [0]{\begingroup\@sanitize@url \@url }%
\providecommand \@url [1]{\endgroup\@href {#1}{\urlprefix }}%
\providecommand \urlprefix  [0]{URL }%
\providecommand \Eprint [0]{\href }%
\providecommand \doibase [0]{https://doi.org/}%
\providecommand \selectlanguage [0]{\@gobble}%
\providecommand \bibinfo  [0]{\@secondoftwo}%
\providecommand \bibfield  [0]{\@secondoftwo}%
\providecommand \translation [1]{[#1]}%
\providecommand \BibitemOpen [0]{}%
\providecommand \bibitemStop [0]{}%
\providecommand \bibitemNoStop [0]{.\EOS\space}%
\providecommand \EOS [0]{\spacefactor3000\relax}%
\providecommand \BibitemShut  [1]{\csname bibitem#1\endcsname}%
\let\auto@bib@innerbib\@empty
\bibitem [{\citenamefont {Quataert}\ and\ \citenamefont {Gruzinov}(1999)}]{Quataert1999-rq}%
  \BibitemOpen
  \bibfield  {author} {\bibinfo {author} {\bibfnamefont {E.}~\bibnamefont {Quataert}}\ and\ \bibinfo {author} {\bibfnamefont {A.}~\bibnamefont {Gruzinov}},\ }\bibfield  {title} {\bibinfo {title} {Turbulence and particle heating in advection‐dominated accretion flows},\ }\href {https://doi.org/10.1086/307423} {\bibfield  {journal} {\bibinfo  {journal} {Astrophys. J.}\ }\textbf {\bibinfo {volume} {520}},\ \bibinfo {pages} {248} (\bibinfo {year} {1999})}\BibitemShut {NoStop}%
\bibitem [{\citenamefont {Ferri{\`e}re}(2001)}]{Ferriere2001}%
  \BibitemOpen
  \bibfield  {author} {\bibinfo {author} {\bibfnamefont {K.~M.}\ \bibnamefont {Ferri{\`e}re}},\ }\bibfield  {title} {\bibinfo {title} {{The interstellar environment of our galaxy}},\ }\href@noop {} {\bibfield  {journal} {\bibinfo  {journal} {Rev. Mod. Phys.}\ }\textbf {\bibinfo {volume} {73}},\ \bibinfo {pages} {1031} (\bibinfo {year} {2001})}\BibitemShut {NoStop}%
\bibitem [{\citenamefont {Kunz}\ \emph {et~al.}(2022)\citenamefont {Kunz}, \citenamefont {Jones},\ and\ \citenamefont {Zhuravleva}}]{Kunz2022}%
  \BibitemOpen
  \bibfield  {author} {\bibinfo {author} {\bibfnamefont {M.~W.}\ \bibnamefont {Kunz}}, \bibinfo {author} {\bibfnamefont {T.~W.}\ \bibnamefont {Jones}},\ and\ \bibinfo {author} {\bibfnamefont {I.}~\bibnamefont {Zhuravleva}},\ }\bibinfo {title} {Plasma physics of the intracluster medium},\ in\ \href {https://doi.org/10.1007/978-981-16-4544-0_125-1} {\emph {\bibinfo {booktitle} {Handbook of X-ray and Gamma-ray Astrophysics}}},\ \bibinfo {editor} {edited by\ \bibinfo {editor} {\bibfnamefont {C.}~\bibnamefont {Bambi}}\ and\ \bibinfo {editor} {\bibfnamefont {A.}~\bibnamefont {Santangelo}}}\ (\bibinfo  {publisher} {Springer},\ \bibinfo {address} {Singapore},\ \bibinfo {year} {2022})\ pp.\ \bibinfo {pages} {1--42}\BibitemShut {NoStop}%
\bibitem [{\citenamefont {Howes}(2024)}]{Howes2024-fn}%
  \BibitemOpen
  \bibfield  {author} {\bibinfo {author} {\bibfnamefont {G.~G.}\ \bibnamefont {Howes}},\ }\bibfield  {title} {\bibinfo {title} {The fundamental parameters of astrophysical plasma turbulence and its dissipation: Nonrelativistic limit},\ }\href {http://arxiv.org/abs/2402.12829} {\bibfield  {journal} {\bibinfo  {journal} {arXiv [astro-ph.SR]}\ } (\bibinfo {year} {2024})},\ \Eprint {https://arxiv.org/abs/2402.12829} {arXiv:2402.12829 [astro-ph.SR]} \BibitemShut {NoStop}%
\bibitem [{\citenamefont {Marsch}\ \emph {et~al.}(1982)\citenamefont {Marsch}, \citenamefont {Mühlhäuser}, \citenamefont {Schwenn}, \citenamefont {Rosenbauer}, \citenamefont {Pilipp},\ and\ \citenamefont {Neubauer}}]{Marsch1982-vu}%
  \BibitemOpen
  \bibfield  {author} {\bibinfo {author} {\bibfnamefont {E.}~\bibnamefont {Marsch}}, \bibinfo {author} {\bibfnamefont {K.-H.}\ \bibnamefont {Mühlhäuser}}, \bibinfo {author} {\bibfnamefont {R.}~\bibnamefont {Schwenn}}, \bibinfo {author} {\bibfnamefont {H.}~\bibnamefont {Rosenbauer}}, \bibinfo {author} {\bibfnamefont {W.}~\bibnamefont {Pilipp}},\ and\ \bibinfo {author} {\bibfnamefont {F.~M.}\ \bibnamefont {Neubauer}},\ }\bibfield  {title} {\bibinfo {title} {Solar wind protons: Three-dimensional velocity distributions and derived plasma parameters measured between 0.3 and 1 {AU}},\ }\href {https://doi.org/10.1029/ja087ia01p00052} {\bibfield  {journal} {\bibinfo  {journal} {J. Geophys. Res.}\ }\textbf {\bibinfo {volume} {87}},\ \bibinfo {pages} {52} (\bibinfo {year} {1982})}\BibitemShut {NoStop}%
\bibitem [{\citenamefont {Marsch}(2004)}]{Marsch2004-fw}%
  \BibitemOpen
  \bibfield  {author} {\bibinfo {author} {\bibfnamefont {E.}~\bibnamefont {Marsch}},\ }\bibfield  {title} {\bibinfo {title} {On the temperature anisotropy of the core part of the proton velocity distribution function in the solar wind},\ }\bibfield  {journal} {\bibinfo  {journal} {J. Geophys. Res.}\ }\textbf {\bibinfo {volume} {109}},\ \href {https://doi.org/10.1029/2003ja010330} {10.1029/2003ja010330} (\bibinfo {year} {2004})\BibitemShut {NoStop}%
\bibitem [{\citenamefont {Hellinger}\ \emph {et~al.}(2006)\citenamefont {Hellinger}, \citenamefont {Trávníček}, \citenamefont {Kasper},\ and\ \citenamefont {Lazarus}}]{Hellinger2006-ff}%
  \BibitemOpen
  \bibfield  {author} {\bibinfo {author} {\bibfnamefont {P.}~\bibnamefont {Hellinger}}, \bibinfo {author} {\bibfnamefont {P.}~\bibnamefont {Trávníček}}, \bibinfo {author} {\bibfnamefont {J.~C.}\ \bibnamefont {Kasper}},\ and\ \bibinfo {author} {\bibfnamefont {A.~J.}\ \bibnamefont {Lazarus}},\ }\bibfield  {title} {\bibinfo {title} {Solar wind proton temperature anisotropy: Linear theory and wind/swe observations},\ }\bibfield  {journal} {\bibinfo  {journal} {Geophys. Res. Lett.}\ }\textbf {\bibinfo {volume} {33}},\ \href {https://doi.org/10.1029/2006gl025925} {10.1029/2006gl025925} (\bibinfo {year} {2006})\BibitemShut {NoStop}%
\bibitem [{\citenamefont {Kohl}\ \emph {et~al.}(1998)\citenamefont {Kohl}, \citenamefont {Noci}, \citenamefont {Antonucci}, \citenamefont {Tondello}, \citenamefont {Huber}, \citenamefont {Cranmer}, \citenamefont {Strachan}, \citenamefont {Panasyuk}, \citenamefont {Gardner}, \citenamefont {Romoli}, \citenamefont {Fineschi}, \citenamefont {Dobrzycka}, \citenamefont {Raymond}, \citenamefont {Nicolosi}, \citenamefont {Siegmund}, \citenamefont {Spadaro}, \citenamefont {Benna}, \citenamefont {Ciaravella}, \citenamefont {Giordano}, \citenamefont {Habbal}, \citenamefont {Karovska}, \citenamefont {Li}, \citenamefont {Martin}, \citenamefont {Michels}, \citenamefont {Modigliani}, \citenamefont {Naletto}, \citenamefont {O'Neal}, \citenamefont {Pernechele}, \citenamefont {Poletto}, \citenamefont {Smith},\ and\ \citenamefont {Suleiman}}]{Kohl1998-cq}%
  \BibitemOpen
  \bibfield  {author} {\bibinfo {author} {\bibfnamefont {J.~L.}\ \bibnamefont {Kohl}}, \bibinfo {author} {\bibfnamefont {G.}~\bibnamefont {Noci}}, \bibinfo {author} {\bibfnamefont {E.}~\bibnamefont {Antonucci}}, \bibinfo {author} {\bibfnamefont {G.}~\bibnamefont {Tondello}}, \bibinfo {author} {\bibfnamefont {M.~C.~E.}\ \bibnamefont {Huber}}, \bibinfo {author} {\bibfnamefont {S.~R.}\ \bibnamefont {Cranmer}}, \bibinfo {author} {\bibfnamefont {L.}~\bibnamefont {Strachan}}, \bibinfo {author} {\bibfnamefont {A.~V.}\ \bibnamefont {Panasyuk}}, \bibinfo {author} {\bibfnamefont {L.~D.}\ \bibnamefont {Gardner}}, \bibinfo {author} {\bibfnamefont {M.}~\bibnamefont {Romoli}}, \bibinfo {author} {\bibfnamefont {S.}~\bibnamefont {Fineschi}}, \bibinfo {author} {\bibfnamefont {D.}~\bibnamefont {Dobrzycka}}, \bibinfo {author} {\bibfnamefont {J.~C.}\ \bibnamefont {Raymond}}, \bibinfo {author} {\bibfnamefont {P.}~\bibnamefont {Nicolosi}}, \bibinfo {author} {\bibfnamefont {O.~H.~W.}\ \bibnamefont {Siegmund}}, \bibinfo {author}
  {\bibfnamefont {D.}~\bibnamefont {Spadaro}}, \bibinfo {author} {\bibfnamefont {C.}~\bibnamefont {Benna}}, \bibinfo {author} {\bibfnamefont {A.}~\bibnamefont {Ciaravella}}, \bibinfo {author} {\bibfnamefont {S.}~\bibnamefont {Giordano}}, \bibinfo {author} {\bibfnamefont {S.~R.}\ \bibnamefont {Habbal}}, \bibinfo {author} {\bibfnamefont {M.}~\bibnamefont {Karovska}}, \bibinfo {author} {\bibfnamefont {X.}~\bibnamefont {Li}}, \bibinfo {author} {\bibfnamefont {R.}~\bibnamefont {Martin}}, \bibinfo {author} {\bibfnamefont {J.~G.}\ \bibnamefont {Michels}}, \bibinfo {author} {\bibfnamefont {A.}~\bibnamefont {Modigliani}}, \bibinfo {author} {\bibfnamefont {G.}~\bibnamefont {Naletto}}, \bibinfo {author} {\bibfnamefont {R.~H.}\ \bibnamefont {O'Neal}}, \bibinfo {author} {\bibfnamefont {C.}~\bibnamefont {Pernechele}}, \bibinfo {author} {\bibfnamefont {G.}~\bibnamefont {Poletto}}, \bibinfo {author} {\bibfnamefont {P.~L.}\ \bibnamefont {Smith}},\ and\ \bibinfo {author} {\bibfnamefont {R.~M.}\ \bibnamefont {Suleiman}},\
  }\bibfield  {title} {\bibinfo {title} {{UVCS/\textit{SOHO}} empirical determinations of anisotropic velocity distributions in the solar corona},\ }\href {https://doi.org/10.1086/311434} {\bibfield  {journal} {\bibinfo  {journal} {Astrophys. J.}\ }\textbf {\bibinfo {volume} {501}},\ \bibinfo {pages} {L127} (\bibinfo {year} {1998})}\BibitemShut {NoStop}%
\bibitem [{\citenamefont {Esser}\ \emph {et~al.}(1999)\citenamefont {Esser}, \citenamefont {Fineschi}, \citenamefont {Dobrzycka}, \citenamefont {Habbal}, \citenamefont {Edgar}, \citenamefont {Raymond}, \citenamefont {Kohl},\ and\ \citenamefont {Guhathakurta}}]{Esser1999-qd}%
  \BibitemOpen
  \bibfield  {author} {\bibinfo {author} {\bibfnamefont {R.}~\bibnamefont {Esser}}, \bibinfo {author} {\bibfnamefont {S.}~\bibnamefont {Fineschi}}, \bibinfo {author} {\bibfnamefont {D.}~\bibnamefont {Dobrzycka}}, \bibinfo {author} {\bibfnamefont {S.~R.}\ \bibnamefont {Habbal}}, \bibinfo {author} {\bibfnamefont {R.~J.}\ \bibnamefont {Edgar}}, \bibinfo {author} {\bibfnamefont {J.~C.}\ \bibnamefont {Raymond}}, \bibinfo {author} {\bibfnamefont {J.~L.}\ \bibnamefont {Kohl}},\ and\ \bibinfo {author} {\bibfnamefont {M.}~\bibnamefont {Guhathakurta}},\ }\bibfield  {title} {\bibinfo {title} {Plasma properties in coronal holes derived from measurements of minor ion spectral lines and polarized white light intensity},\ }\href {https://doi.org/10.1086/311786} {\bibfield  {journal} {\bibinfo  {journal} {Astrophys. J.}\ }\textbf {\bibinfo {volume} {510}},\ \bibinfo {pages} {L63} (\bibinfo {year} {1999})}\BibitemShut {NoStop}%
\bibitem [{\citenamefont {Antonucci}(2000)}]{Antonucci2000-lz}%
  \BibitemOpen
  \bibfield  {author} {\bibinfo {author} {\bibfnamefont {E.}~\bibnamefont {Antonucci}},\ }\bibfield  {title} {\bibinfo {title} {Fast solar wind velocity in a polar coronal hole during solar minimum},\ }\href {https://doi.org/10.1023/a:1026568912809} {\bibfield  {journal} {\bibinfo  {journal} {Solar Phys.}\ }\textbf {\bibinfo {volume} {197}},\ \bibinfo {pages} {115} (\bibinfo {year} {2000})}\BibitemShut {NoStop}%
\bibitem [{\citenamefont {Cranmer}\ \emph {et~al.}(2009)\citenamefont {Cranmer}, \citenamefont {Matthaeus}, \citenamefont {Breech},\ and\ \citenamefont {Kasper}}]{Cranmer2009-vf}%
  \BibitemOpen
  \bibfield  {author} {\bibinfo {author} {\bibfnamefont {S.~R.}\ \bibnamefont {Cranmer}}, \bibinfo {author} {\bibfnamefont {W.~H.}\ \bibnamefont {Matthaeus}}, \bibinfo {author} {\bibfnamefont {B.~A.}\ \bibnamefont {Breech}},\ and\ \bibinfo {author} {\bibfnamefont {J.~C.}\ \bibnamefont {Kasper}},\ }\bibfield  {title} {\bibinfo {title} {Empirical constraints on proton and electron heating in the fast solar wind},\ }\href {https://doi.org/10.1088/0004-637x/702/2/1604} {\bibfield  {journal} {\bibinfo  {journal} {Astrophys. J.}\ }\textbf {\bibinfo {volume} {702}},\ \bibinfo {pages} {1604} (\bibinfo {year} {2009})}\BibitemShut {NoStop}%
\bibitem [{\citenamefont {Chandran}\ \emph {et~al.}(2010)\citenamefont {Chandran}, \citenamefont {Li}, \citenamefont {Rogers}, \citenamefont {Quataert},\ and\ \citenamefont {Germaschewski}}]{Chandran2010-ow}%
  \BibitemOpen
  \bibfield  {author} {\bibinfo {author} {\bibfnamefont {B.~D.~G.}\ \bibnamefont {Chandran}}, \bibinfo {author} {\bibfnamefont {B.}~\bibnamefont {Li}}, \bibinfo {author} {\bibfnamefont {B.~N.}\ \bibnamefont {Rogers}}, \bibinfo {author} {\bibfnamefont {E.}~\bibnamefont {Quataert}},\ and\ \bibinfo {author} {\bibfnamefont {K.}~\bibnamefont {Germaschewski}},\ }\bibfield  {title} {\bibinfo {title} {Perpendicular ion heating by low-frequency {Alfvén}-wave turbulence in the solar wind},\ }\href {https://doi.org/10.1088/0004-637X/720/1/503} {\bibfield  {journal} {\bibinfo  {journal} {Astrophys. J.}\ }\textbf {\bibinfo {volume} {720}},\ \bibinfo {pages} {503} (\bibinfo {year} {2010})}\BibitemShut {NoStop}%
\bibitem [{\citenamefont {Chandran}\ \emph {et~al.}(2013)\citenamefont {Chandran}, \citenamefont {Verscharen}, \citenamefont {Quataert}, \citenamefont {Kasper}, \citenamefont {Isenberg},\ and\ \citenamefont {Bourouaine}}]{Chandran2013-ng}%
  \BibitemOpen
  \bibfield  {author} {\bibinfo {author} {\bibfnamefont {B.~D.~G.}\ \bibnamefont {Chandran}}, \bibinfo {author} {\bibfnamefont {D.}~\bibnamefont {Verscharen}}, \bibinfo {author} {\bibfnamefont {E.}~\bibnamefont {Quataert}}, \bibinfo {author} {\bibfnamefont {J.~C.}\ \bibnamefont {Kasper}}, \bibinfo {author} {\bibfnamefont {P.~A.}\ \bibnamefont {Isenberg}},\ and\ \bibinfo {author} {\bibfnamefont {S.}~\bibnamefont {Bourouaine}},\ }\bibfield  {title} {\bibinfo {title} {Stochastic heating, differential flow, and the alpha-to-proton temperature ratio in the solar wind},\ }\href {https://doi.org/10.1088/0004-637X/776/1/45} {\bibfield  {journal} {\bibinfo  {journal} {ApJ}\ }\textbf {\bibinfo {volume} {776}},\ \bibinfo {pages} {45} (\bibinfo {year} {2013})}\BibitemShut {NoStop}%
\bibitem [{\citenamefont {Hollweg}\ and\ \citenamefont {Isenberg}(2002)}]{Hollweg2002-dw}%
  \BibitemOpen
  \bibfield  {author} {\bibinfo {author} {\bibfnamefont {J.~V.}\ \bibnamefont {Hollweg}}\ and\ \bibinfo {author} {\bibfnamefont {P.~A.}\ \bibnamefont {Isenberg}},\ }\bibfield  {title} {\bibinfo {title} {Generation of the fast solar wind: A review with emphasis on the resonant cyclotron interaction},\ }\href {https://doi.org/10.1029/2001JA000270} {\bibfield  {journal} {\bibinfo  {journal} {Journal of Geophysical Research: Space Physics}\ }\textbf {\bibinfo {volume} {107}},\ \bibinfo {pages} {SSH 12} (\bibinfo {year} {2002})}\BibitemShut {NoStop}%
\bibitem [{\citenamefont {Isenberg}\ and\ \citenamefont {Vasquez}(2011)}]{Isenberg2011-wt}%
  \BibitemOpen
  \bibfield  {author} {\bibinfo {author} {\bibfnamefont {P.~A.}\ \bibnamefont {Isenberg}}\ and\ \bibinfo {author} {\bibfnamefont {B.~J.}\ \bibnamefont {Vasquez}},\ }\bibfield  {title} {\bibinfo {title} {A kinetic model of solar wind generation by oblique ion-cyclotron waves},\ }\href {https://doi.org/10.1088/0004-637X/731/2/88} {\bibfield  {journal} {\bibinfo  {journal} {ApJ}\ }\textbf {\bibinfo {volume} {731}},\ \bibinfo {pages} {88} (\bibinfo {year} {2011})}\BibitemShut {NoStop}%
\bibitem [{\citenamefont {Isenberg}\ and\ \citenamefont {Vasquez}(2019)}]{Isenberg2019-oy}%
  \BibitemOpen
  \bibfield  {author} {\bibinfo {author} {\bibfnamefont {P.~A.}\ \bibnamefont {Isenberg}}\ and\ \bibinfo {author} {\bibfnamefont {B.~J.}\ \bibnamefont {Vasquez}},\ }\bibfield  {title} {\bibinfo {title} {Perpendicular ion heating by cyclotron resonant dissipation of turbulently generated kinetic {Alfvén} waves in the solar wind},\ }\href {https://doi.org/10.3847/1538-4357/ab4e12} {\bibfield  {journal} {\bibinfo  {journal} {Astrophys. J.}\ }\textbf {\bibinfo {volume} {887}},\ \bibinfo {pages} {63} (\bibinfo {year} {2019})}\BibitemShut {NoStop}%
\bibitem [{\citenamefont {Chandran}(2005)}]{Chandran2005-mb}%
  \BibitemOpen
  \bibfield  {author} {\bibinfo {author} {\bibfnamefont {B.~D.~G.}\ \bibnamefont {Chandran}},\ }\bibfield  {title} {\bibinfo {title} {Weak compressible magnetohydrodynamic turbulence in the solar corona},\ }\href {https://doi.org/10.1103/PhysRevLett.95.265004} {\bibfield  {journal} {\bibinfo  {journal} {Phys. Rev. Lett.}\ }\textbf {\bibinfo {volume} {95}},\ \bibinfo {pages} {265004} (\bibinfo {year} {2005})}\BibitemShut {NoStop}%
\bibitem [{\citenamefont {Schekochihin}\ \emph {et~al.}(2009)\citenamefont {Schekochihin}, \citenamefont {Cowley}, \citenamefont {Dorland}, \citenamefont {Hammett}, \citenamefont {Howes}, \citenamefont {Quataert},\ and\ \citenamefont {Tatsuno}}]{Schekochihin2009-qo}%
  \BibitemOpen
  \bibfield  {author} {\bibinfo {author} {\bibfnamefont {A.~A.}\ \bibnamefont {Schekochihin}}, \bibinfo {author} {\bibfnamefont {S.~C.}\ \bibnamefont {Cowley}}, \bibinfo {author} {\bibfnamefont {W.}~\bibnamefont {Dorland}}, \bibinfo {author} {\bibfnamefont {G.~W.}\ \bibnamefont {Hammett}}, \bibinfo {author} {\bibfnamefont {G.~G.}\ \bibnamefont {Howes}}, \bibinfo {author} {\bibfnamefont {E.}~\bibnamefont {Quataert}},\ and\ \bibinfo {author} {\bibfnamefont {T.}~\bibnamefont {Tatsuno}},\ }\bibfield  {title} {\bibinfo {title} {Astrophysical gyrokinetics: kinetic and fluid turbulent cascades in magnetized weakly collisional plasmas},\ }\href {https://doi.org/10.1088/0067-0049/182/1/310} {\bibfield  {journal} {\bibinfo  {journal} {Astrophys. J. Supp.}\ }\textbf {\bibinfo {volume} {182}},\ \bibinfo {pages} {310} (\bibinfo {year} {2009})}\BibitemShut {NoStop}%
\bibitem [{\citenamefont {TenBarge}\ and\ \citenamefont {Howes}(2013)}]{TenBarge2013-ux}%
  \BibitemOpen
  \bibfield  {author} {\bibinfo {author} {\bibfnamefont {J.~M.}\ \bibnamefont {TenBarge}}\ and\ \bibinfo {author} {\bibfnamefont {G.~G.}\ \bibnamefont {Howes}},\ }\bibfield  {title} {\bibinfo {title} {Current sheets and collisionless damping in kinetic plasma turbulence},\ }\href {https://doi.org/10.1088/2041-8205/771/2/L27} {\bibfield  {journal} {\bibinfo  {journal} {ApJL}\ }\textbf {\bibinfo {volume} {771}},\ \bibinfo {pages} {L27} (\bibinfo {year} {2013})}\BibitemShut {NoStop}%
\bibitem [{\citenamefont {Lehe}\ \emph {et~al.}(2009)\citenamefont {Lehe}, \citenamefont {Parrish},\ and\ \citenamefont {Quataert}}]{Lehe2009-up}%
  \BibitemOpen
  \bibfield  {author} {\bibinfo {author} {\bibfnamefont {R.}~\bibnamefont {Lehe}}, \bibinfo {author} {\bibfnamefont {I.~J.}\ \bibnamefont {Parrish}},\ and\ \bibinfo {author} {\bibfnamefont {E.}~\bibnamefont {Quataert}},\ }\bibfield  {title} {\bibinfo {title} {The heating of test particles in numerical simulations of {Alfvénic} turbulence},\ }\href {https://doi.org/10.1088/0004-637X/707/1/404} {\bibfield  {journal} {\bibinfo  {journal} {Astrophys. J.}\ }\textbf {\bibinfo {volume} {707}},\ \bibinfo {pages} {404} (\bibinfo {year} {2009})}\BibitemShut {NoStop}%
\bibitem [{\citenamefont {Xia}\ \emph {et~al.}(2013)\citenamefont {Xia}, \citenamefont {Perez}, \citenamefont {Chandran},\ and\ \citenamefont {Quataert}}]{Xia2013-ob}%
  \BibitemOpen
  \bibfield  {author} {\bibinfo {author} {\bibfnamefont {Q.}~\bibnamefont {Xia}}, \bibinfo {author} {\bibfnamefont {J.~C.}\ \bibnamefont {Perez}}, \bibinfo {author} {\bibfnamefont {B.~D.~G.}\ \bibnamefont {Chandran}},\ and\ \bibinfo {author} {\bibfnamefont {E.}~\bibnamefont {Quataert}},\ }\bibfield  {title} {\bibinfo {title} {Perpendicular ion heating by reduced magnetohydrodynamic turbulence},\ }\href {https://doi.org/10.1088/0004-637X/776/2/90} {\bibfield  {journal} {\bibinfo  {journal} {Astrophys. J.}\ }\textbf {\bibinfo {volume} {776}},\ \bibinfo {pages} {90} (\bibinfo {year} {2013})}\BibitemShut {NoStop}%
\bibitem [{\citenamefont {Teaca}\ \emph {et~al.}(2014)\citenamefont {Teaca}, \citenamefont {Weidl}, \citenamefont {Jenko},\ and\ \citenamefont {Schlickeiser}}]{Teaca2014-iq}%
  \BibitemOpen
  \bibfield  {author} {\bibinfo {author} {\bibfnamefont {B.}~\bibnamefont {Teaca}}, \bibinfo {author} {\bibfnamefont {M.~S.}\ \bibnamefont {Weidl}}, \bibinfo {author} {\bibfnamefont {F.}~\bibnamefont {Jenko}},\ and\ \bibinfo {author} {\bibfnamefont {R.}~\bibnamefont {Schlickeiser}},\ }\bibfield  {title} {\bibinfo {title} {Acceleration of particles in imbalanced magnetohydrodynamic turbulence},\ }\href {https://doi.org/10.1103/PhysRevE.90.021101} {\bibfield  {journal} {\bibinfo  {journal} {Phys. Rev. E Stat. Nonlin. Soft Matter Phys.}\ }\textbf {\bibinfo {volume} {90}},\ \bibinfo {pages} {021101} (\bibinfo {year} {2014})}\BibitemShut {NoStop}%
\bibitem [{\citenamefont {Weidl}\ \emph {et~al.}(2015)\citenamefont {Weidl}, \citenamefont {Jenko}, \citenamefont {Teaca},\ and\ \citenamefont {Schlickeiser}}]{Weidl2015-sd}%
  \BibitemOpen
  \bibfield  {author} {\bibinfo {author} {\bibfnamefont {M.~S.}\ \bibnamefont {Weidl}}, \bibinfo {author} {\bibfnamefont {F.}~\bibnamefont {Jenko}}, \bibinfo {author} {\bibfnamefont {B.}~\bibnamefont {Teaca}},\ and\ \bibinfo {author} {\bibfnamefont {R.}~\bibnamefont {Schlickeiser}},\ }\bibfield  {title} {\bibinfo {title} {Cosmic-ray pitch-angle scattering in imbalanced {MHD} turbulence simulations},\ }\href {https://doi.org/10.1088/0004-637X/811/1/8} {\bibfield  {journal} {\bibinfo  {journal} {ApJ}\ }\textbf {\bibinfo {volume} {811}},\ \bibinfo {pages} {8} (\bibinfo {year} {2015})}\BibitemShut {NoStop}%
\bibitem [{\citenamefont {Arzamasskiy}\ \emph {et~al.}(2019)\citenamefont {Arzamasskiy}, \citenamefont {Kunz}, \citenamefont {Chandran},\ and\ \citenamefont {Quataert}}]{Arzamasskiy2019-qv}%
  \BibitemOpen
  \bibfield  {author} {\bibinfo {author} {\bibfnamefont {L.}~\bibnamefont {Arzamasskiy}}, \bibinfo {author} {\bibfnamefont {M.~W.}\ \bibnamefont {Kunz}}, \bibinfo {author} {\bibfnamefont {B.~D.~G.}\ \bibnamefont {Chandran}},\ and\ \bibinfo {author} {\bibfnamefont {E.}~\bibnamefont {Quataert}},\ }\bibfield  {title} {\bibinfo {title} {Hybrid-kinetic simulations of ion heating in {Alfvénic} turbulence},\ }\href {https://doi.org/10.3847/1538-4357/ab20cc} {\bibfield  {journal} {\bibinfo  {journal} {Astrophys. J.}\ }\textbf {\bibinfo {volume} {879}},\ \bibinfo {pages} {53} (\bibinfo {year} {2019})}\BibitemShut {NoStop}%
\bibitem [{\citenamefont {Cerri}\ \emph {et~al.}(2021)\citenamefont {Cerri}, \citenamefont {Arzamasskiy},\ and\ \citenamefont {Kunz}}]{Cerri2021-xo}%
  \BibitemOpen
  \bibfield  {author} {\bibinfo {author} {\bibfnamefont {S.~S.}\ \bibnamefont {Cerri}}, \bibinfo {author} {\bibfnamefont {L.}~\bibnamefont {Arzamasskiy}},\ and\ \bibinfo {author} {\bibfnamefont {M.~W.}\ \bibnamefont {Kunz}},\ }\bibfield  {title} {\bibinfo {title} {On stochastic heating and its phase-space signatures in low-beta kinetic turbulence},\ }\href {https://doi.org/10.3847/1538-4357/abfbde} {\bibfield  {journal} {\bibinfo  {journal} {Astrophys. J.}\ }\textbf {\bibinfo {volume} {916}},\ \bibinfo {pages} {120} (\bibinfo {year} {2021})}\BibitemShut {NoStop}%
\bibitem [{\citenamefont {Pugliese}\ \emph {et~al.}(2023)\citenamefont {Pugliese}, \citenamefont {Brodiano}, \citenamefont {Andrés},\ and\ \citenamefont {Dmitruk}}]{Pugliese2023-px}%
  \BibitemOpen
  \bibfield  {author} {\bibinfo {author} {\bibfnamefont {F.}~\bibnamefont {Pugliese}}, \bibinfo {author} {\bibfnamefont {M.}~\bibnamefont {Brodiano}}, \bibinfo {author} {\bibfnamefont {N.}~\bibnamefont {Andrés}},\ and\ \bibinfo {author} {\bibfnamefont {P.}~\bibnamefont {Dmitruk}},\ }\bibfield  {title} {\bibinfo {title} {Energization of charged test particles in magnetohydrodynamic fields: Waves versus turbulence picture},\ }\href {https://doi.org/10.3847/1538-4357/ad055b} {\bibfield  {journal} {\bibinfo  {journal} {Astrophys. J.}\ }\textbf {\bibinfo {volume} {959}},\ \bibinfo {pages} {28} (\bibinfo {year} {2023})}\BibitemShut {NoStop}%
\bibitem [{\citenamefont {Bourouaine}\ and\ \citenamefont {Chandran}(2013)}]{Bourouaine2013-gh}%
  \BibitemOpen
  \bibfield  {author} {\bibinfo {author} {\bibfnamefont {S.}~\bibnamefont {Bourouaine}}\ and\ \bibinfo {author} {\bibfnamefont {B.~D.~G.}\ \bibnamefont {Chandran}},\ }\bibfield  {title} {\bibinfo {title} {Observational test of stochastic heating in low-$\beta$ fast-solar-wind streams},\ }\href {https://doi.org/10.1088/0004-637X/774/2/96} {\bibfield  {journal} {\bibinfo  {journal} {ApJ}\ }\textbf {\bibinfo {volume} {774}},\ \bibinfo {pages} {96} (\bibinfo {year} {2013})}\BibitemShut {NoStop}%
\bibitem [{\citenamefont {Klein}\ and\ \citenamefont {Chandran}(2016)}]{Klein2016-qo}%
  \BibitemOpen
  \bibfield  {author} {\bibinfo {author} {\bibfnamefont {K.~G.}\ \bibnamefont {Klein}}\ and\ \bibinfo {author} {\bibfnamefont {B.~D.~G.}\ \bibnamefont {Chandran}},\ }\bibfield  {title} {\bibinfo {title} {Evolution of the proton velocity distribution due to stochastic heating in the near-sun solar wind},\ }\href {https://doi.org/10.3847/0004-637X/820/1/47} {\bibfield  {journal} {\bibinfo  {journal} {Astrophys. J.}\ }\textbf {\bibinfo {volume} {820}},\ \bibinfo {pages} {47} (\bibinfo {year} {2016})}\BibitemShut {NoStop}%
\bibitem [{\citenamefont {Vech}\ \emph {et~al.}(2017)\citenamefont {Vech}, \citenamefont {Klein},\ and\ \citenamefont {Kasper}}]{Vech2017-jq}%
  \BibitemOpen
  \bibfield  {author} {\bibinfo {author} {\bibfnamefont {D.}~\bibnamefont {Vech}}, \bibinfo {author} {\bibfnamefont {K.~G.}\ \bibnamefont {Klein}},\ and\ \bibinfo {author} {\bibfnamefont {J.~C.}\ \bibnamefont {Kasper}},\ }\bibfield  {title} {\bibinfo {title} {Nature of stochastic ion heating in the solar wind: Testing the dependence on plasma beta and turbulence amplitude},\ }\href {https://doi.org/10.3847/2041-8213/aa9887} {\bibfield  {journal} {\bibinfo  {journal} {Astrophys. J. Lett.}\ }\textbf {\bibinfo {volume} {850}},\ \bibinfo {pages} {L11} (\bibinfo {year} {2017})}\BibitemShut {NoStop}%
\bibitem [{\citenamefont {Martinović}\ \emph {et~al.}(2019)\citenamefont {Martinović}, \citenamefont {Klein},\ and\ \citenamefont {Bourouaine}}]{Martinovic2019-xf}%
  \BibitemOpen
  \bibfield  {author} {\bibinfo {author} {\bibfnamefont {M.~M.}\ \bibnamefont {Martinović}}, \bibinfo {author} {\bibfnamefont {K.~G.}\ \bibnamefont {Klein}},\ and\ \bibinfo {author} {\bibfnamefont {S.}~\bibnamefont {Bourouaine}},\ }\bibfield  {title} {\bibinfo {title} {Radial evolution of stochastic heating in low-$\beta$ solar wind},\ }\href {https://doi.org/10.3847/1538-4357/ab23f4} {\bibfield  {journal} {\bibinfo  {journal} {ApJ}\ }\textbf {\bibinfo {volume} {879}},\ \bibinfo {pages} {43} (\bibinfo {year} {2019})}\BibitemShut {NoStop}%
\bibitem [{\citenamefont {Martinović}\ \emph {et~al.}(2020)\citenamefont {Martinović}, \citenamefont {Klein}, \citenamefont {Kasper}, \citenamefont {Case}, \citenamefont {Korreck}, \citenamefont {Larson}, \citenamefont {Livi}, \citenamefont {Stevens}, \citenamefont {Whittlesey}, \citenamefont {Chandran}, \citenamefont {Alterman}, \citenamefont {Huang}, \citenamefont {Chen}, \citenamefont {Bale}, \citenamefont {Pulupa}, \citenamefont {Malaspina}, \citenamefont {Bonnell}, \citenamefont {Harvey}, \citenamefont {Goetz}, \citenamefont {Dudok~de Wit},\ and\ \citenamefont {MacDowall}}]{Martinovic2020-do}%
  \BibitemOpen
  \bibfield  {author} {\bibinfo {author} {\bibfnamefont {M.~M.}\ \bibnamefont {Martinović}}, \bibinfo {author} {\bibfnamefont {K.~G.}\ \bibnamefont {Klein}}, \bibinfo {author} {\bibfnamefont {J.~C.}\ \bibnamefont {Kasper}}, \bibinfo {author} {\bibfnamefont {A.~W.}\ \bibnamefont {Case}}, \bibinfo {author} {\bibfnamefont {K.~E.}\ \bibnamefont {Korreck}}, \bibinfo {author} {\bibfnamefont {D.}~\bibnamefont {Larson}}, \bibinfo {author} {\bibfnamefont {R.}~\bibnamefont {Livi}}, \bibinfo {author} {\bibfnamefont {M.}~\bibnamefont {Stevens}}, \bibinfo {author} {\bibfnamefont {P.}~\bibnamefont {Whittlesey}}, \bibinfo {author} {\bibfnamefont {B.~D.~G.}\ \bibnamefont {Chandran}}, \bibinfo {author} {\bibfnamefont {B.~L.}\ \bibnamefont {Alterman}}, \bibinfo {author} {\bibfnamefont {J.}~\bibnamefont {Huang}}, \bibinfo {author} {\bibfnamefont {C.~H.~K.}\ \bibnamefont {Chen}}, \bibinfo {author} {\bibfnamefont {S.~D.}\ \bibnamefont {Bale}}, \bibinfo {author} {\bibfnamefont {M.}~\bibnamefont {Pulupa}}, \bibinfo {author}
  {\bibfnamefont {D.~M.}\ \bibnamefont {Malaspina}}, \bibinfo {author} {\bibfnamefont {J.~W.}\ \bibnamefont {Bonnell}}, \bibinfo {author} {\bibfnamefont {P.~R.}\ \bibnamefont {Harvey}}, \bibinfo {author} {\bibfnamefont {K.}~\bibnamefont {Goetz}}, \bibinfo {author} {\bibfnamefont {T.}~\bibnamefont {Dudok~de Wit}},\ and\ \bibinfo {author} {\bibfnamefont {R.~J.}\ \bibnamefont {MacDowall}},\ }\bibfield  {title} {\bibinfo {title} {The enhancement of proton stochastic heating in the near-{Sun} solar wind},\ }\href {https://doi.org/10.3847/1538-4365/ab527f} {\bibfield  {journal} {\bibinfo  {journal} {Astrophys. J. Suppl. Ser.}\ }\textbf {\bibinfo {volume} {246}},\ \bibinfo {pages} {30} (\bibinfo {year} {2020})}\BibitemShut {NoStop}%
\bibitem [{\citenamefont {McChesney}\ \emph {et~al.}(1987)\citenamefont {McChesney}, \citenamefont {Stern},\ and\ \citenamefont {Bellan}}]{McChesney1987-dp}%
  \BibitemOpen
  \bibfield  {author} {\bibinfo {author} {\bibfnamefont {J.~M.}\ \bibnamefont {McChesney}}, \bibinfo {author} {\bibfnamefont {R.~A.}\ \bibnamefont {Stern}},\ and\ \bibinfo {author} {\bibfnamefont {P.~M.}\ \bibnamefont {Bellan}},\ }\bibfield  {title} {\bibinfo {title} {Observation of fast stochastic ion heating by drift waves},\ }\href {https://doi.org/10.1103/PhysRevLett.59.1436} {\bibfield  {journal} {\bibinfo  {journal} {Phys. Rev. Lett.}\ }\textbf {\bibinfo {volume} {59}},\ \bibinfo {pages} {1436} (\bibinfo {year} {1987})}\BibitemShut {NoStop}%
\bibitem [{\citenamefont {Johnson}\ and\ \citenamefont {Cheng}(2001)}]{Johnson2001-ui}%
  \BibitemOpen
  \bibfield  {author} {\bibinfo {author} {\bibfnamefont {J.~R.}\ \bibnamefont {Johnson}}\ and\ \bibinfo {author} {\bibfnamefont {C.~Z.}\ \bibnamefont {Cheng}},\ }\bibfield  {title} {\bibinfo {title} {Stochastic ion heating at the magnetopause due to kinetic {Alfvén} waves},\ }\href {https://doi.org/10.1029/2001gl013509} {\bibfield  {journal} {\bibinfo  {journal} {Geophys. Res. Lett.}\ }\textbf {\bibinfo {volume} {28}},\ \bibinfo {pages} {4421} (\bibinfo {year} {2001})}\BibitemShut {NoStop}%
\bibitem [{\citenamefont {Kennel}\ and\ \citenamefont {Engelmann}(1966)}]{Kennel1966-rx}%
  \BibitemOpen
  \bibfield  {author} {\bibinfo {author} {\bibfnamefont {C.~F.}\ \bibnamefont {Kennel}}\ and\ \bibinfo {author} {\bibfnamefont {F.}~\bibnamefont {Engelmann}},\ }\bibfield  {title} {\bibinfo {title} {Velocity space diffusion from weak plasma turbulence in a magnetic field},\ }\href {https://doi.org/10.1063/1.1761629} {\bibfield  {journal} {\bibinfo  {journal} {Phys. Fluids}\ }\textbf {\bibinfo {volume} {9}},\ \bibinfo {pages} {2377} (\bibinfo {year} {1966})}\BibitemShut {NoStop}%
\bibitem [{\citenamefont {Stix}(1992)}]{Stix1992-bo}%
  \BibitemOpen
  \bibfield  {author} {\bibinfo {author} {\bibfnamefont {T.~H.}\ \bibnamefont {Stix}},\ }\href@noop {} {\emph {\bibinfo {title} {Waves in Plasmas}}},\ \bibinfo {edition} {1992nd}\ ed.\ (\bibinfo  {publisher} {American Institute of Physics},\ \bibinfo {address} {New York, NY},\ \bibinfo {year} {1992})\BibitemShut {NoStop}%
\bibitem [{\citenamefont {Schlickeiser}\ and\ \citenamefont {Achatz}(1993)}]{Schlickeiser1993-if}%
  \BibitemOpen
  \bibfield  {author} {\bibinfo {author} {\bibfnamefont {R.}~\bibnamefont {Schlickeiser}}\ and\ \bibinfo {author} {\bibfnamefont {U.}~\bibnamefont {Achatz}},\ }\bibfield  {title} {\bibinfo {title} {Cosmic-ray particle transport in weakly turbulent plasmas. part 1. theory},\ }\href {https://doi.org/10.1017/s0022377800016822} {\bibfield  {journal} {\bibinfo  {journal} {J. Plasma Phys.}\ }\textbf {\bibinfo {volume} {49}},\ \bibinfo {pages} {63} (\bibinfo {year} {1993})}\BibitemShut {NoStop}%
\bibitem [{\citenamefont {Hoppock}\ \emph {et~al.}(2018)\citenamefont {Hoppock}, \citenamefont {Chandran}, \citenamefont {Klein}, \citenamefont {Mallet},\ and\ \citenamefont {Verscharen}}]{Hoppock2018-fv}%
  \BibitemOpen
  \bibfield  {author} {\bibinfo {author} {\bibfnamefont {I.~W.}\ \bibnamefont {Hoppock}}, \bibinfo {author} {\bibfnamefont {B.~D.~G.}\ \bibnamefont {Chandran}}, \bibinfo {author} {\bibfnamefont {K.~G.}\ \bibnamefont {Klein}}, \bibinfo {author} {\bibfnamefont {A.}~\bibnamefont {Mallet}},\ and\ \bibinfo {author} {\bibfnamefont {D.}~\bibnamefont {Verscharen}},\ }\bibfield  {title} {\bibinfo {title} {Stochastic proton heating by kinetic-{Alfvén}-wave turbulence in moderately high-$\beta$ plasmas},\ }\bibfield  {journal} {\bibinfo  {journal} {J. Plasma Phys.}\ }\textbf {\bibinfo {volume} {84}},\ \href {https://doi.org/10.1017/S0022377818001277} {10.1017/S0022377818001277} (\bibinfo {year} {2018})\BibitemShut {NoStop}%
\bibitem [{\citenamefont {Zhang}\ \emph {et~al.}(2025)\citenamefont {Zhang}, \citenamefont {Kunz}, \citenamefont {Squire},\ and\ \citenamefont {Klein}}]{Zhang2025-yd}%
  \BibitemOpen
  \bibfield  {author} {\bibinfo {author} {\bibfnamefont {M.~F.}\ \bibnamefont {Zhang}}, \bibinfo {author} {\bibfnamefont {M.~W.}\ \bibnamefont {Kunz}}, \bibinfo {author} {\bibfnamefont {J.}~\bibnamefont {Squire}},\ and\ \bibinfo {author} {\bibfnamefont {K.~G.}\ \bibnamefont {Klein}},\ }\bibfield  {title} {\bibinfo {title} {Extreme heating of minor ions in imbalanced solar-wind turbulence},\ }\href {https://doi.org/10.3847/1538-4357/ad95fc} {\bibfield  {journal} {\bibinfo  {journal} {Astrophys. J.}\ }\textbf {\bibinfo {volume} {979}},\ \bibinfo {pages} {121} (\bibinfo {year} {2025})}\BibitemShut {NoStop}%
\bibitem [{\citenamefont {Bowen}\ \emph {et~al.}(2022)\citenamefont {Bowen}, \citenamefont {Chandran}, \citenamefont {Squire}, \citenamefont {Bale}, \citenamefont {Duan}, \citenamefont {Klein}, \citenamefont {Larson}, \citenamefont {Mallet}, \citenamefont {McManus}, \citenamefont {Meyrand}, \citenamefont {Verniero},\ and\ \citenamefont {Woodham}}]{Bowen2022-qt}%
  \BibitemOpen
  \bibfield  {author} {\bibinfo {author} {\bibfnamefont {T.~A.}\ \bibnamefont {Bowen}}, \bibinfo {author} {\bibfnamefont {B.~D.~G.}\ \bibnamefont {Chandran}}, \bibinfo {author} {\bibfnamefont {J.}~\bibnamefont {Squire}}, \bibinfo {author} {\bibfnamefont {S.~D.}\ \bibnamefont {Bale}}, \bibinfo {author} {\bibfnamefont {D.}~\bibnamefont {Duan}}, \bibinfo {author} {\bibfnamefont {K.~G.}\ \bibnamefont {Klein}}, \bibinfo {author} {\bibfnamefont {D.}~\bibnamefont {Larson}}, \bibinfo {author} {\bibfnamefont {A.}~\bibnamefont {Mallet}}, \bibinfo {author} {\bibfnamefont {M.~D.}\ \bibnamefont {McManus}}, \bibinfo {author} {\bibfnamefont {R.}~\bibnamefont {Meyrand}}, \bibinfo {author} {\bibfnamefont {J.~L.}\ \bibnamefont {Verniero}},\ and\ \bibinfo {author} {\bibfnamefont {L.~D.}\ \bibnamefont {Woodham}},\ }\bibfield  {title} {\bibinfo {title} {\textit{In situ} signature of cyclotron resonant heating in the solar wind},\ }\href {https://doi.org/10.1103/PhysRevLett.129.165101} {\bibfield  {journal} {\bibinfo  {journal}
  {Phys. Rev. Lett.}\ }\textbf {\bibinfo {volume} {129}},\ \bibinfo {pages} {165101} (\bibinfo {year} {2022})}\BibitemShut {NoStop}%
\bibitem [{\citenamefont {Bowen}\ \emph {et~al.}(2024)\citenamefont {Bowen}, \citenamefont {Bale}, \citenamefont {Chandran}, \citenamefont {Chasapis}, \citenamefont {Chen}, \citenamefont {Dudok~de Wit}, \citenamefont {Mallet}, \citenamefont {Meyrand},\ and\ \citenamefont {Squire}}]{Bowen2024-si}%
  \BibitemOpen
  \bibfield  {author} {\bibinfo {author} {\bibfnamefont {T.~A.}\ \bibnamefont {Bowen}}, \bibinfo {author} {\bibfnamefont {S.~D.}\ \bibnamefont {Bale}}, \bibinfo {author} {\bibfnamefont {B.~D.~G.}\ \bibnamefont {Chandran}}, \bibinfo {author} {\bibfnamefont {A.}~\bibnamefont {Chasapis}}, \bibinfo {author} {\bibfnamefont {C.~H.~K.}\ \bibnamefont {Chen}}, \bibinfo {author} {\bibfnamefont {T.}~\bibnamefont {Dudok~de Wit}}, \bibinfo {author} {\bibfnamefont {A.}~\bibnamefont {Mallet}}, \bibinfo {author} {\bibfnamefont {R.}~\bibnamefont {Meyrand}},\ and\ \bibinfo {author} {\bibfnamefont {J.}~\bibnamefont {Squire}},\ }\bibfield  {title} {\bibinfo {title} {Mediation of collisionless turbulent dissipation through cyclotron resonance},\ }\href {https://doi.org/10.1038/s41550-023-02186-4} {\bibfield  {journal} {\bibinfo  {journal} {Nat. Astron.}\ ,\ \bibinfo {pages} {1}} (\bibinfo {year} {2024})}\BibitemShut {NoStop}%
\bibitem [{\citenamefont {Schekochihin}(2022)}]{Schekochihin2022-nn}%
  \BibitemOpen
  \bibfield  {author} {\bibinfo {author} {\bibfnamefont {A.~A.}\ \bibnamefont {Schekochihin}},\ }\bibfield  {title} {\bibinfo {title} {{MHD} turbulence: a biased review},\ }\href {https://doi.org/10.1017/S0022377822000721} {\bibfield  {journal} {\bibinfo  {journal} {J. Plasma Phys.}\ }\textbf {\bibinfo {volume} {88}},\ \bibinfo {pages} {155880501} (\bibinfo {year} {2022})}\BibitemShut {NoStop}%
\bibitem [{\citenamefont {Goldreich}\ and\ \citenamefont {Sridhar}(1995)}]{Goldreich1995-fv}%
  \BibitemOpen
  \bibfield  {author} {\bibinfo {author} {\bibfnamefont {P.}~\bibnamefont {Goldreich}}\ and\ \bibinfo {author} {\bibfnamefont {S.}~\bibnamefont {Sridhar}},\ }\bibfield  {title} {\bibinfo {title} {Toward a theory of interstellar turbulence. 2: Strong {Alfvénic} turbulence},\ }\href {https://doi.org/10.1086/175121} {\bibfield  {journal} {\bibinfo  {journal} {Astrophys. J.}\ }\textbf {\bibinfo {volume} {438}},\ \bibinfo {pages} {763} (\bibinfo {year} {1995})}\BibitemShut {NoStop}%
\bibitem [{\citenamefont {Meyrand}\ \emph {et~al.}(2021)\citenamefont {Meyrand}, \citenamefont {Squire}, \citenamefont {Schekochihin},\ and\ \citenamefont {Dorland}}]{Meyrand2021-ix}%
  \BibitemOpen
  \bibfield  {author} {\bibinfo {author} {\bibfnamefont {R.}~\bibnamefont {Meyrand}}, \bibinfo {author} {\bibfnamefont {J.}~\bibnamefont {Squire}}, \bibinfo {author} {\bibfnamefont {A.~A.}\ \bibnamefont {Schekochihin}},\ and\ \bibinfo {author} {\bibfnamefont {W.}~\bibnamefont {Dorland}},\ }\bibfield  {title} {\bibinfo {title} {On the violation of the zeroth law of turbulence in space plasmas},\ }\bibfield  {journal} {\bibinfo  {journal} {J. Plasma Phys.}\ }\textbf {\bibinfo {volume} {87}},\ \href {https://doi.org/10.1017/S0022377821000489} {10.1017/S0022377821000489} (\bibinfo {year} {2021})\BibitemShut {NoStop}%
\bibitem [{\citenamefont {Squire}\ \emph {et~al.}(2022)\citenamefont {Squire}, \citenamefont {Meyrand}, \citenamefont {Kunz}, \citenamefont {Arzamasskiy}, \citenamefont {Schekochihin},\ and\ \citenamefont {Quataert}}]{Squire2022-dm}%
  \BibitemOpen
  \bibfield  {author} {\bibinfo {author} {\bibfnamefont {J.}~\bibnamefont {Squire}}, \bibinfo {author} {\bibfnamefont {R.}~\bibnamefont {Meyrand}}, \bibinfo {author} {\bibfnamefont {M.~W.}\ \bibnamefont {Kunz}}, \bibinfo {author} {\bibfnamefont {L.}~\bibnamefont {Arzamasskiy}}, \bibinfo {author} {\bibfnamefont {A.~A.}\ \bibnamefont {Schekochihin}},\ and\ \bibinfo {author} {\bibfnamefont {E.}~\bibnamefont {Quataert}},\ }\bibfield  {title} {\bibinfo {title} {High-frequency heating of the solar wind triggered by low-frequency turbulence},\ }\href {https://doi.org/10.1038/s41550-022-01624-z} {\bibfield  {journal} {\bibinfo  {journal} {Nature Astronomy}\ }\textbf {\bibinfo {volume} {6}},\ \bibinfo {pages} {715} (\bibinfo {year} {2022})}\BibitemShut {NoStop}%
\bibitem [{\citenamefont {Squire}\ \emph {et~al.}(2023)\citenamefont {Squire}, \citenamefont {Meyrand},\ and\ \citenamefont {Kunz}}]{Squire2023-jn}%
  \BibitemOpen
  \bibfield  {author} {\bibinfo {author} {\bibfnamefont {J.}~\bibnamefont {Squire}}, \bibinfo {author} {\bibfnamefont {R.}~\bibnamefont {Meyrand}},\ and\ \bibinfo {author} {\bibfnamefont {M.~W.}\ \bibnamefont {Kunz}},\ }\bibfield  {title} {\bibinfo {title} {Electron–ion heating partition in imbalanced solar-wind turbulence},\ }\href {https://doi.org/10.3847/2041-8213/ad0779} {\bibfield  {journal} {\bibinfo  {journal} {Astrophys. J. Lett.}\ }\textbf {\bibinfo {volume} {957}},\ \bibinfo {pages} {L30} (\bibinfo {year} {2023})}\BibitemShut {NoStop}%
\bibitem [{\citenamefont {Lithwick}\ \emph {et~al.}(2007)\citenamefont {Lithwick}, \citenamefont {Goldreich},\ and\ \citenamefont {Sridhar}}]{Lithwick2007-ao}%
  \BibitemOpen
  \bibfield  {author} {\bibinfo {author} {\bibfnamefont {Y.}~\bibnamefont {Lithwick}}, \bibinfo {author} {\bibfnamefont {P.}~\bibnamefont {Goldreich}},\ and\ \bibinfo {author} {\bibfnamefont {S.}~\bibnamefont {Sridhar}},\ }\bibfield  {title} {\bibinfo {title} {Imbalanced strong {MHD} turbulence},\ }\href {https://doi.org/10.1086/509884} {\bibfield  {journal} {\bibinfo  {journal} {Astrophys. J.}\ }\textbf {\bibinfo {volume} {655}},\ \bibinfo {pages} {269} (\bibinfo {year} {2007})}\BibitemShut {NoStop}%
\bibitem [{\citenamefont {Beresnyak}\ and\ \citenamefont {Lazarian}(2008)}]{Beresnyak2008-aw}%
  \BibitemOpen
  \bibfield  {author} {\bibinfo {author} {\bibfnamefont {A.}~\bibnamefont {Beresnyak}}\ and\ \bibinfo {author} {\bibfnamefont {A.}~\bibnamefont {Lazarian}},\ }\bibfield  {title} {\bibinfo {title} {Strong imbalanced turbulence},\ }\href {https://doi.org/10.1086/589428} {\bibfield  {journal} {\bibinfo  {journal} {Astrophys. J.}\ }\textbf {\bibinfo {volume} {682}},\ \bibinfo {pages} {1070} (\bibinfo {year} {2008})}\BibitemShut {NoStop}%
\bibitem [{\citenamefont {Chandran}(2000)}]{Chandran2000-pq}%
  \BibitemOpen
  \bibfield  {author} {\bibinfo {author} {\bibfnamefont {B.~D.}\ \bibnamefont {Chandran}},\ }\bibfield  {title} {\bibinfo {title} {Scattering of energetic particles by anisotropic magnetohydrodynamic turbulence with a {Goldreich-Sridhar} power spectrum},\ }\href {https://doi.org/10.1103/PhysRevLett.85.4656} {\bibfield  {journal} {\bibinfo  {journal} {Phys. Rev. Lett.}\ }\textbf {\bibinfo {volume} {85}},\ \bibinfo {pages} {4656} (\bibinfo {year} {2000})}\BibitemShut {NoStop}%
\bibitem [{\citenamefont {Passot}\ \emph {et~al.}(2018)\citenamefont {Passot}, \citenamefont {Sulem},\ and\ \citenamefont {Tassi}}]{Passot2018-yt}%
  \BibitemOpen
  \bibfield  {author} {\bibinfo {author} {\bibfnamefont {T.}~\bibnamefont {Passot}}, \bibinfo {author} {\bibfnamefont {P.~L.}\ \bibnamefont {Sulem}},\ and\ \bibinfo {author} {\bibfnamefont {E.}~\bibnamefont {Tassi}},\ }\bibfield  {title} {\bibinfo {title} {Gyrofluid modeling and phenomenology of low-$\beta_e$\textit{ }{Alfvén} wave turbulence},\ }\href {https://doi.org/10.1063/1.5022528} {\bibfield  {journal} {\bibinfo  {journal} {Phys. Plasmas}\ }\textbf {\bibinfo {volume} {25}},\ \bibinfo {pages} {042107} (\bibinfo {year} {2018})}\BibitemShut {NoStop}%
\bibitem [{\citenamefont {Schekochihin}\ \emph {et~al.}(2019)\citenamefont {Schekochihin}, \citenamefont {Kawazura},\ and\ \citenamefont {Barnes}}]{Schekochihin2019-al}%
  \BibitemOpen
  \bibfield  {author} {\bibinfo {author} {\bibfnamefont {A.~A.}\ \bibnamefont {Schekochihin}}, \bibinfo {author} {\bibfnamefont {Y.}~\bibnamefont {Kawazura}},\ and\ \bibinfo {author} {\bibfnamefont {M.~A.}\ \bibnamefont {Barnes}},\ }\bibfield  {title} {\bibinfo {title} {Constraints on ion versus electron heating by plasma turbulence at low beta},\ }\bibfield  {journal} {\bibinfo  {journal} {J. Plasma Phys.}\ }\textbf {\bibinfo {volume} {85}},\ \href {https://doi.org/10.1017/S0022377819000345} {10.1017/S0022377819000345} (\bibinfo {year} {2019})\BibitemShut {NoStop}%
\bibitem [{\citenamefont {Strauss}(1976)}]{Strauss1976-vu}%
  \BibitemOpen
  \bibfield  {author} {\bibinfo {author} {\bibfnamefont {H.~R.}\ \bibnamefont {Strauss}},\ }\bibfield  {title} {\bibinfo {title} {Nonlinear, three-dimensional magnetohydrodynamics of noncircular tokamaks},\ }\href {https://doi.org/10.1063/1.861310} {\bibfield  {journal} {\bibinfo  {journal} {Phys. Fluids}\ }\textbf {\bibinfo {volume} {19}},\ \bibinfo {pages} {134} (\bibinfo {year} {1976})}\BibitemShut {NoStop}%
\bibitem [{\citenamefont {Boldyrev}\ \emph {et~al.}(2013)\citenamefont {Boldyrev}, \citenamefont {Horaites}, \citenamefont {Xia},\ and\ \citenamefont {Perez}}]{Boldyrev2013-gq}%
  \BibitemOpen
  \bibfield  {author} {\bibinfo {author} {\bibfnamefont {S.}~\bibnamefont {Boldyrev}}, \bibinfo {author} {\bibfnamefont {K.}~\bibnamefont {Horaites}}, \bibinfo {author} {\bibfnamefont {Q.}~\bibnamefont {Xia}},\ and\ \bibinfo {author} {\bibfnamefont {J.~C.}\ \bibnamefont {Perez}},\ }\bibfield  {title} {\bibinfo {title} {Toward a theory of astrophysical plasma turbulence at subproton scales},\ }\href {https://doi.org/10.1088/0004-637x/777/1/41} {\bibfield  {journal} {\bibinfo  {journal} {Astrophys. J.}\ }\textbf {\bibinfo {volume} {777}},\ \bibinfo {pages} {41} (\bibinfo {year} {2013})}\BibitemShut {NoStop}%
\bibitem [{\citenamefont {Teaca}\ \emph {et~al.}(2009)\citenamefont {Teaca}, \citenamefont {Verma}, \citenamefont {Knaepen},\ and\ \citenamefont {Carati}}]{Teaca2009-go}%
  \BibitemOpen
  \bibfield  {author} {\bibinfo {author} {\bibfnamefont {B.}~\bibnamefont {Teaca}}, \bibinfo {author} {\bibfnamefont {M.~K.}\ \bibnamefont {Verma}}, \bibinfo {author} {\bibfnamefont {B.}~\bibnamefont {Knaepen}},\ and\ \bibinfo {author} {\bibfnamefont {D.}~\bibnamefont {Carati}},\ }\bibfield  {title} {\bibinfo {title} {Energy transfer in anisotropic magnetohydrodynamic turbulence},\ }\href {https://doi.org/10.1103/PhysRevE.79.046312} {\bibfield  {journal} {\bibinfo  {journal} {Phys. Rev. E Stat. Nonlin. Soft Matter Phys.}\ }\textbf {\bibinfo {volume} {79}},\ \bibinfo {pages} {046312} (\bibinfo {year} {2009})}\BibitemShut {NoStop}%
\bibitem [{\citenamefont {Bott}\ \emph {et~al.}(2021)\citenamefont {Bott}, \citenamefont {Arzamasskiy}, \citenamefont {Kunz}, \citenamefont {Quataert},\ and\ \citenamefont {Squire}}]{Bott2021-vp}%
  \BibitemOpen
  \bibfield  {author} {\bibinfo {author} {\bibfnamefont {A.~F.~A.}\ \bibnamefont {Bott}}, \bibinfo {author} {\bibfnamefont {L.}~\bibnamefont {Arzamasskiy}}, \bibinfo {author} {\bibfnamefont {M.~W.}\ \bibnamefont {Kunz}}, \bibinfo {author} {\bibfnamefont {E.}~\bibnamefont {Quataert}},\ and\ \bibinfo {author} {\bibfnamefont {J.}~\bibnamefont {Squire}},\ }\bibfield  {title} {\bibinfo {title} {Adaptive critical balance and firehose instability in an expanding, turbulent, collisionless plasma},\ }\href {https://doi.org/10.3847/2041-8213/ac37c2} {\bibfield  {journal} {\bibinfo  {journal} {Astrophys. J. Lett.}\ }\textbf {\bibinfo {volume} {922}},\ \bibinfo {pages} {L35} (\bibinfo {year} {2021})}\BibitemShut {NoStop}%
\bibitem [{\citenamefont {Kiyani}\ \emph {et~al.}(2015)\citenamefont {Kiyani}, \citenamefont {Osman},\ and\ \citenamefont {Chapman}}]{Kiyani2015-yk}%
  \BibitemOpen
  \bibfield  {author} {\bibinfo {author} {\bibfnamefont {K.~H.}\ \bibnamefont {Kiyani}}, \bibinfo {author} {\bibfnamefont {K.~T.}\ \bibnamefont {Osman}},\ and\ \bibinfo {author} {\bibfnamefont {S.~C.}\ \bibnamefont {Chapman}},\ }\bibfield  {title} {\bibinfo {title} {Dissipation and heating in solar wind turbulence: from the macro to the micro and back again},\ }\bibfield  {journal} {\bibinfo  {journal} {Philos. Trans. A Math. Phys. Eng. Sci.}\ }\textbf {\bibinfo {volume} {373}},\ \href {https://doi.org/10.1098/rsta.2014.0155} {10.1098/rsta.2014.0155} (\bibinfo {year} {2015})\BibitemShut {NoStop}%
\bibitem [{\citenamefont {Bowen}\ \emph {et~al.}(2020)\citenamefont {Bowen}, \citenamefont {Mallet}, \citenamefont {Bale}, \citenamefont {Bonnell}, \citenamefont {Case}, \citenamefont {Chandran}, \citenamefont {Chasapis}, \citenamefont {Chen}, \citenamefont {Duan}, \citenamefont {Dudok~de Wit}, \citenamefont {Goetz}, \citenamefont {Halekas}, \citenamefont {Harvey}, \citenamefont {Kasper}, \citenamefont {Korreck}, \citenamefont {Larson}, \citenamefont {Livi}, \citenamefont {MacDowall}, \citenamefont {Malaspina}, \citenamefont {McManus}, \citenamefont {Pulupa}, \citenamefont {Stevens},\ and\ \citenamefont {Whittlesey}}]{Bowen2020-wa}%
  \BibitemOpen
  \bibfield  {author} {\bibinfo {author} {\bibfnamefont {T.~A.}\ \bibnamefont {Bowen}}, \bibinfo {author} {\bibfnamefont {A.}~\bibnamefont {Mallet}}, \bibinfo {author} {\bibfnamefont {S.~D.}\ \bibnamefont {Bale}}, \bibinfo {author} {\bibfnamefont {J.~W.}\ \bibnamefont {Bonnell}}, \bibinfo {author} {\bibfnamefont {A.~W.}\ \bibnamefont {Case}}, \bibinfo {author} {\bibfnamefont {B.~D.~G.}\ \bibnamefont {Chandran}}, \bibinfo {author} {\bibfnamefont {A.}~\bibnamefont {Chasapis}}, \bibinfo {author} {\bibfnamefont {C.~H.~K.}\ \bibnamefont {Chen}}, \bibinfo {author} {\bibfnamefont {D.}~\bibnamefont {Duan}}, \bibinfo {author} {\bibfnamefont {T.}~\bibnamefont {Dudok~de Wit}}, \bibinfo {author} {\bibfnamefont {K.}~\bibnamefont {Goetz}}, \bibinfo {author} {\bibfnamefont {J.~S.}\ \bibnamefont {Halekas}}, \bibinfo {author} {\bibfnamefont {P.~R.}\ \bibnamefont {Harvey}}, \bibinfo {author} {\bibfnamefont {J.~C.}\ \bibnamefont {Kasper}}, \bibinfo {author} {\bibfnamefont {K.~E.}\ \bibnamefont {Korreck}}, \bibinfo {author}
  {\bibfnamefont {D.}~\bibnamefont {Larson}}, \bibinfo {author} {\bibfnamefont {R.}~\bibnamefont {Livi}}, \bibinfo {author} {\bibfnamefont {R.~J.}\ \bibnamefont {MacDowall}}, \bibinfo {author} {\bibfnamefont {D.~M.}\ \bibnamefont {Malaspina}}, \bibinfo {author} {\bibfnamefont {M.~D.}\ \bibnamefont {McManus}}, \bibinfo {author} {\bibfnamefont {M.}~\bibnamefont {Pulupa}}, \bibinfo {author} {\bibfnamefont {M.}~\bibnamefont {Stevens}},\ and\ \bibinfo {author} {\bibfnamefont {P.}~\bibnamefont {Whittlesey}},\ }\bibfield  {title} {\bibinfo {title} {Constraining ion-scale heating and spectral energy transfer in observations of plasma turbulence},\ }\href {https://doi.org/10.1103/PhysRevLett.125.025102} {\bibfield  {journal} {\bibinfo  {journal} {Phys. Rev. Lett.}\ }\textbf {\bibinfo {volume} {125}},\ \bibinfo {pages} {025102} (\bibinfo {year} {2020})}\BibitemShut {NoStop}%
\bibitem [{\citenamefont {Adkins}\ \emph {et~al.}(2024)\citenamefont {Adkins}, \citenamefont {Meyrand},\ and\ \citenamefont {Squire}}]{Adkins2024-xf}%
  \BibitemOpen
  \bibfield  {author} {\bibinfo {author} {\bibfnamefont {T.}~\bibnamefont {Adkins}}, \bibinfo {author} {\bibfnamefont {R.}~\bibnamefont {Meyrand}},\ and\ \bibinfo {author} {\bibfnamefont {J.}~\bibnamefont {Squire}},\ }\bibfield  {title} {\bibinfo {title} {The effects of finite electron inertia on helicity-barrier-mediated turbulence},\ }\href {http://arxiv.org/abs/2404.09380} {\bibfield  {journal} {\bibinfo  {journal} {arXiv [physics.plasm-ph]}\ } (\bibinfo {year} {2024})},\ \Eprint {https://arxiv.org/abs/2404.09380} {arXiv:2404.09380 [physics.plasm-ph]} \BibitemShut {NoStop}%
\bibitem [{\citenamefont {Leamon}\ \emph {et~al.}(1998)\citenamefont {Leamon}, \citenamefont {Smith}, \citenamefont {Ness}, \citenamefont {Matthaeus},\ and\ \citenamefont {Wong}}]{Leamon1998-kl}%
  \BibitemOpen
  \bibfield  {author} {\bibinfo {author} {\bibfnamefont {R.~J.}\ \bibnamefont {Leamon}}, \bibinfo {author} {\bibfnamefont {C.~W.}\ \bibnamefont {Smith}}, \bibinfo {author} {\bibfnamefont {N.~F.}\ \bibnamefont {Ness}}, \bibinfo {author} {\bibfnamefont {W.~H.}\ \bibnamefont {Matthaeus}},\ and\ \bibinfo {author} {\bibfnamefont {H.~K.}\ \bibnamefont {Wong}},\ }\bibfield  {title} {\bibinfo {title} {Observational constraints on the dynamics of the interplanetary magnetic field dissipation range},\ }\href {https://doi.org/10.1029/97JA03394} {\bibfield  {journal} {\bibinfo  {journal} {Journal of Geophysical Research: Space Physics}\ }\textbf {\bibinfo {volume} {103}},\ \bibinfo {pages} {4775} (\bibinfo {year} {1998})}\BibitemShut {NoStop}%
\bibitem [{\citenamefont {McIntyre}\ \emph {et~al.}(2024)\citenamefont {McIntyre}, \citenamefont {Chen}, \citenamefont {Squire}, \citenamefont {Meyrand},\ and\ \citenamefont {Simon}}]{McIntyre2024-pb}%
  \BibitemOpen
  \bibfield  {author} {\bibinfo {author} {\bibfnamefont {J.~R.}\ \bibnamefont {McIntyre}}, \bibinfo {author} {\bibfnamefont {C.~H.~K.}\ \bibnamefont {Chen}}, \bibinfo {author} {\bibfnamefont {J.}~\bibnamefont {Squire}}, \bibinfo {author} {\bibfnamefont {R.}~\bibnamefont {Meyrand}},\ and\ \bibinfo {author} {\bibfnamefont {P.~A.}\ \bibnamefont {Simon}},\ }\bibfield  {title} {\bibinfo {title} {Evidence for the helicity barrier from measurements of the turbulence transition range in the solar wind},\ }\href {http://arxiv.org/abs/2407.10815} {\bibfield  {journal} {\bibinfo  {journal} {arXiv [astro-ph.SR]}\ } (\bibinfo {year} {2024})},\ \Eprint {https://arxiv.org/abs/2407.10815} {arXiv:2407.10815 [astro-ph.SR]} \BibitemShut {NoStop}%
\bibitem [{\citenamefont {Boldyrev}(2006)}]{Boldyrev2006-xd}%
  \BibitemOpen
  \bibfield  {author} {\bibinfo {author} {\bibfnamefont {S.}~\bibnamefont {Boldyrev}},\ }\bibfield  {title} {\bibinfo {title} {Spectrum of magnetohydrodynamic turbulence},\ }\href {https://doi.org/10.1103/PhysRevLett.96.115002} {\bibfield  {journal} {\bibinfo  {journal} {Phys. Rev. Lett.}\ }\textbf {\bibinfo {volume} {96}},\ \bibinfo {pages} {115002} (\bibinfo {year} {2006})}\BibitemShut {NoStop}%
\bibitem [{\citenamefont {Howes}\ \emph {et~al.}(2006)\citenamefont {Howes}, \citenamefont {Cowley}, \citenamefont {Dorland}, \citenamefont {Hammett}, \citenamefont {Quataert},\ and\ \citenamefont {Schekochihin}}]{Howes2006-tq}%
  \BibitemOpen
  \bibfield  {author} {\bibinfo {author} {\bibfnamefont {G.~G.}\ \bibnamefont {Howes}}, \bibinfo {author} {\bibfnamefont {S.~C.}\ \bibnamefont {Cowley}}, \bibinfo {author} {\bibfnamefont {W.}~\bibnamefont {Dorland}}, \bibinfo {author} {\bibfnamefont {G.~W.}\ \bibnamefont {Hammett}}, \bibinfo {author} {\bibfnamefont {E.}~\bibnamefont {Quataert}},\ and\ \bibinfo {author} {\bibfnamefont {A.~A.}\ \bibnamefont {Schekochihin}},\ }\bibfield  {title} {\bibinfo {title} {Astrophysical gyrokinetics: Basic equations and linear theory},\ }\href {https://doi.org/10.1086/506172} {\bibfield  {journal} {\bibinfo  {journal} {Astrophys. J.}\ }\textbf {\bibinfo {volume} {651}},\ \bibinfo {pages} {590} (\bibinfo {year} {2006})}\BibitemShut {NoStop}%
\bibitem [{\citenamefont {Zocco}\ and\ \citenamefont {Schekochihin}(2011)}]{Zocco2011-zp}%
  \BibitemOpen
  \bibfield  {author} {\bibinfo {author} {\bibfnamefont {A.}~\bibnamefont {Zocco}}\ and\ \bibinfo {author} {\bibfnamefont {A.~A.}\ \bibnamefont {Schekochihin}},\ }\bibfield  {title} {\bibinfo {title} {Reduced fluid-kinetic equations for low-frequency dynamics, magnetic reconnection, and electron heating in low-beta plasmas},\ }\href {https://doi.org/10.1063/1.3628639} {\bibfield  {journal} {\bibinfo  {journal} {Phys. Plasmas}\ }\textbf {\bibinfo {volume} {18}},\ \bibinfo {pages} {102309} (\bibinfo {year} {2011})}\BibitemShut {NoStop}%
\bibitem [{\citenamefont {Meyrand}\ \emph {et~al.}(2019)\citenamefont {Meyrand}, \citenamefont {Kanekar}, \citenamefont {Dorland},\ and\ \citenamefont {Schekochihin}}]{Meyrand2019-ms}%
  \BibitemOpen
  \bibfield  {author} {\bibinfo {author} {\bibfnamefont {R.}~\bibnamefont {Meyrand}}, \bibinfo {author} {\bibfnamefont {A.}~\bibnamefont {Kanekar}}, \bibinfo {author} {\bibfnamefont {W.}~\bibnamefont {Dorland}},\ and\ \bibinfo {author} {\bibfnamefont {A.~A.}\ \bibnamefont {Schekochihin}},\ }\bibfield  {title} {\bibinfo {title} {Fluidization of collisionless plasma turbulence},\ }\href {https://doi.org/10.1073/pnas.1813913116} {\bibfield  {journal} {\bibinfo  {journal} {Proc. Natl. Acad. Sci. U. S. A.}\ }\textbf {\bibinfo {volume} {116}},\ \bibinfo {pages} {1185} (\bibinfo {year} {2019})}\BibitemShut {NoStop}%
\bibitem [{\citenamefont {David}\ \emph {et~al.}(2024)\citenamefont {David}, \citenamefont {Galtier},\ and\ \citenamefont {Meyrand}}]{David2024-bm}%
  \BibitemOpen
  \bibfield  {author} {\bibinfo {author} {\bibfnamefont {V.}~\bibnamefont {David}}, \bibinfo {author} {\bibfnamefont {S.}~\bibnamefont {Galtier}},\ and\ \bibinfo {author} {\bibfnamefont {R.}~\bibnamefont {Meyrand}},\ }\bibfield  {title} {\bibinfo {title} {Monofractality in the solar wind at electron scales: Insights from kinetic {Alfvén} waves turbulence},\ }\href {https://doi.org/10.1103/physrevlett.132.085201} {\bibfield  {journal} {\bibinfo  {journal} {Phys. Rev. Lett.}\ }\textbf {\bibinfo {volume} {132}},\ \bibinfo {pages} {085201} (\bibinfo {year} {2024})}\BibitemShut {NoStop}%
\bibitem [{\citenamefont {Williamson}(1980)}]{Williamson1980-hz}%
  \BibitemOpen
  \bibfield  {author} {\bibinfo {author} {\bibfnamefont {J.~H.}\ \bibnamefont {Williamson}},\ }\bibfield  {title} {\bibinfo {title} {Low-storage {Runge-Kutta} schemes},\ }\href {https://doi.org/10.1016/0021-9991(80)90033-9} {\bibfield  {journal} {\bibinfo  {journal} {J. Comput. Phys.}\ }\textbf {\bibinfo {volume} {35}},\ \bibinfo {pages} {48} (\bibinfo {year} {1980})}\BibitemShut {NoStop}%
\bibitem [{\citenamefont {Mallet}\ \emph {et~al.}(2017)\citenamefont {Mallet}, \citenamefont {Schekochihin},\ and\ \citenamefont {Chandran}}]{Mallet2017-mg}%
  \BibitemOpen
  \bibfield  {author} {\bibinfo {author} {\bibfnamefont {A.}~\bibnamefont {Mallet}}, \bibinfo {author} {\bibfnamefont {A.~A.}\ \bibnamefont {Schekochihin}},\ and\ \bibinfo {author} {\bibfnamefont {B.~D.~G.}\ \bibnamefont {Chandran}},\ }\bibfield  {title} {\bibinfo {title} {Disruption of {Alfvénic} turbulence by magnetic reconnection in a collisionless plasma},\ }\href {https://doi.org/10.1017/s0022377817000812} {\bibfield  {journal} {\bibinfo  {journal} {J. Plasma Phys.}\ }\textbf {\bibinfo {volume} {83}},\ \bibinfo {pages} {905830609} (\bibinfo {year} {2017})}\BibitemShut {NoStop}%
\bibitem [{\citenamefont {Loureiro}\ and\ \citenamefont {Boldyrev}(2017)}]{Loureiro2017-xo}%
  \BibitemOpen
  \bibfield  {author} {\bibinfo {author} {\bibfnamefont {N.~F.}\ \bibnamefont {Loureiro}}\ and\ \bibinfo {author} {\bibfnamefont {S.}~\bibnamefont {Boldyrev}},\ }\bibfield  {title} {\bibinfo {title} {Collisionless reconnection in magnetohydrodynamic and kinetic turbulence},\ }\href {https://doi.org/10.3847/1538-4357/aa9754} {\bibfield  {journal} {\bibinfo  {journal} {Astrophys. J.}\ }\textbf {\bibinfo {volume} {850}},\ \bibinfo {pages} {182} (\bibinfo {year} {2017})}\BibitemShut {NoStop}%
\bibitem [{\citenamefont {Zhou}\ \emph {et~al.}(2023)\citenamefont {Zhou}, \citenamefont {Liu},\ and\ \citenamefont {Loureiro}}]{Zhou2023-vm}%
  \BibitemOpen
  \bibfield  {author} {\bibinfo {author} {\bibfnamefont {M.}~\bibnamefont {Zhou}}, \bibinfo {author} {\bibfnamefont {Z.}~\bibnamefont {Liu}},\ and\ \bibinfo {author} {\bibfnamefont {N.~F.}\ \bibnamefont {Loureiro}},\ }\bibfield  {title} {\bibinfo {title} {Electron heating in kinetic-{Alfvén}-wave turbulence},\ }\href {https://doi.org/10.1073/pnas.2220927120} {\bibfield  {journal} {\bibinfo  {journal} {Proc. Natl. Acad. Sci. U. S. A.}\ }\textbf {\bibinfo {volume} {120}},\ \bibinfo {pages} {e2220927120} (\bibinfo {year} {2023})}\BibitemShut {NoStop}%
\bibitem [{\citenamefont {Boris}(1970)}]{Boris1970-sf}%
  \BibitemOpen
  \bibfield  {author} {\bibinfo {author} {\bibfnamefont {J.~P.}\ \bibnamefont {Boris}},\ }\bibfield  {title} {\bibinfo {title} {Relativistic plasma simulation-optimization},\ }in\ \href@noop {} {\emph {\bibinfo {booktitle} {Proc. Fourth Conf. Num. Sim. Plasmas}}}\ (\bibinfo {year} {1970})\ pp.\ \bibinfo {pages} {3--67}\BibitemShut {NoStop}%
\bibitem [{\citenamefont {Birdsall}\ and\ \citenamefont {Langdon}(2018)}]{Birdsall2018-mv}%
  \BibitemOpen
  \bibfield  {author} {\bibinfo {author} {\bibfnamefont {C.~K.}\ \bibnamefont {Birdsall}}\ and\ \bibinfo {author} {\bibfnamefont {A.~B.}\ \bibnamefont {Langdon}},\ }\href {https://www.taylorfrancis.com/books/mono/10.1201/9781315275048/plasma-physics-via-computer-simulation-birdsall-langdon} {\emph {\bibinfo {title} {Plasma physics via computer simulation}}}\ (\bibinfo  {publisher} {CRC Press},\ \bibinfo {year} {2018})\BibitemShut {NoStop}%
\end{thebibliography}%

\section*{End Matter}

\textit{The Helicity Barrier.}---Here we describe the basics physics of the helicity barrier, as referenced in the Letter, which necessarily accompanies imbalanced turbulence in a collisionless plasma.
While it does not directly control the frequency spectrum of the turbulence, its influence on heating is felt via the creation of the sharp ``transition range" drop in the spectrum around $\ion{\rho}$.

The helicity barrier occurs because low-$\beta$ gyrokinetics conserves energy and a ``generalized helicity" \cite{Schekochihin2019-al,Meyrand2021-ix}, which must be simultaneously cascaded with constant flux to be in a statistical steady-state.
At scales larger than $\ion{\rho}$, helicity is the cross-helicity which cascades to small perpendicular scales.
At scales smaller than $\ion{\rho}$, helicity becomes magnetic helicity, which would inverse cascade in imbalanced turbulence above electron-skin-depth scales \cite{Adkins2024-xf}.
For balanced turbulence, with zero helicity, energy is able to cascade to small perpendicular scales.
However, in imbalanced turbulence only the balanced (zero-helicity) portion of the energy cascade is let through to small perpendicular scales; the remainder is blocked from cascading and trapped at scales above $\ion{\rho}$ by the helicity barrier.
This causes the amplitude of the turbulence to grow in time and, through critical balance \cite{Goldreich1995-fv}, reach and dissipate at small parallel scales.
Hybrid-kinetic imbalanced turbulence simulations \cite{Squire2022-dm} show that fluctuations at these scales become oblique ion-cyclotron waves, heating ions through quasi-linear interactions and allowing the turbulence to saturate.
The helicity barrier is well supported observationally, explaining the steep ``transition-range" drop in fluctuation spectra \cite{Leamon1998-kl,Bowen2020-wa} and other features \cite{Bowen2022-qt,Bowen2024-si,McIntyre2024-pb}.

\begin{figure}
\includegraphics[scale=0.7]{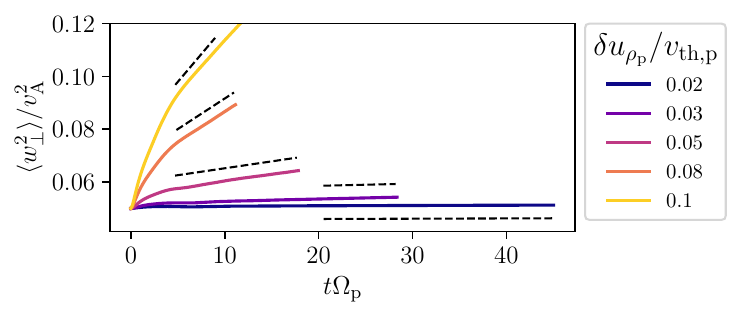}
\caption{\label{fig:wprp_evo} Evolution of the ensemble-averaged perpendicular energy per unit mass $\langle w^2_\perp\rangle$ of all particle cases with $0.02 \leq \deltaup/\vthp \leq 0.1$ from \flrimbalsix, along with the linear fit for $Q_\perp$ in dashed lines.}
\end{figure}

\textit{Measurement of Test Particle $Q_\perp$.}---Figure \ref{fig:wprp_evo} illustrates how $Q_\perp$ is calculated from the ensemble-averaged perpendicular energy per unit mass $\langle w^2_\perp\rangle$ for protons from \flrimbalsix~with $0.02 \leq \deltaup/\vthp \leq 0.1$.
Here, $w_\perp$ is the component perpendicular to $\bm{B}$ of $\bm{w} \equiv \bm{v}-\bm{u}_{\ExB}$, the proton's thermal velocity after subtracting the local $\ExB$ velocity.
We calculate $Q_\perp$ by a linear fit via $Q_\perp = 0.5(\langle w^2_{\perp\text{,f}}\rangle - \langle w^2_{\perp\text{,}0}\rangle)/(t_{\text{f}}-t_0)$, with $t_0$ and $t_f$ representing the start and end times of the fit and corresponding subscripts on $\langle w^2_\perp\rangle$ representing its value at these times.
Ions experience an initial heating transient as they ``pick up" the local $\ExB$ velocity (also seen in Ref.~\cite{Xia2013-ob}); for $\deltaui/\vthi\lesssim 0.05$, these transients become oscillatory.
To account for this, we choose the start time $t_0\Omega_{\rm i}/2\pi = 5.25$ for $\deltaui/\vthi < 0.02$, $3.25$ for $0.02 \leq \deltaui/\vthi \leq 0.05$, and $0.75$ for $\deltaui/\vthi > 0.05$ (see Supplemental Material).
The end time $t_{\text{f}}$ is the minimum of the end of the simulation or when $\langle w^2_{\perp\text{,f}}\rangle = 1.2\langle w^2_{\perp\text{,}0}\rangle$; we choose this to calculate the correct heating rate as $Q_\perp$ decreases with increasing $\langle w^2_\perp\rangle$ \cite{Chandran2010-ow,Xia2013-ob}.

\textit{Ion Heating from a Fully Imbalanced Spectrum of Alfv\'en Waves}---
The form of the parameter $\xi_i$ used in the Letter, as it applies to the helicity barrier, can be motivated by using the machinery of quasi-linear theory \cite{Kennel1966-rx}.
For simplicity, assume fully imbalanced Alfv\'enic turbulence with a two-dimensional RMHD spectra \cite{Schekochihin2022-nn} cut off at perpendicular scales $\lambda\gtrsim\ion{\rho}$ to mimic the effect of the helicity barrier:
\begin{eqnarray}\label{eq:RMHDSpecNorm}
    &E_{\rm 2D}(k_\perp,k_\|)=V\vA^2\ion{\rho}^2\sqrt{\beta}\times\nonumber\\
    &\begin{cases}
        \left(\ion{\rho}/\lambda\right)^{1/3}(\deltaulambda/\vthi)\kprptilde^{-7/3}, & |\kprltilde|\leq\kprltilde^{\rm CB}\\
        \left(\ion{\rho}/\lambda\right)^{2}\left(\deltaulambda/\vthi\right)^{6}|\kprltilde|^{-5}\kprptilde, & |\kprltilde|\geq \kprltilde^{\rm CB}\\ 
        0, & \kprptilde > \ion{\rho}/\lambda \text{ or } |\kprltilde| > 1,
    \end{cases}
\end{eqnarray}
with normalized wavenumbers $\kprltilde\equiv k_\|\vA/\ion{\Omega}$ and $\kprptilde\equiv k_\perp\ion{\rho}$, $\kprltilde^{\rm CB} \equiv \left(\ion{\rho}/\lambda\right)^{1/3}\left(\deltaulambda/\vthi\right)\kprptilde^{2/3}$, and the spectrum is normalized such that
\begin{equation*}
    \frac{1}{\vthi^2}\int_{e^{-1/2}/\lambda}^{e^{1/2}/\lambda} dk_\perp \int_{-\infty}^{\infty}d k_\| E_{\rm 2D}(k_\perp,k_\|)\approx \left(\frac{\deltaulambda}{\vthi}\right)^2.
\end{equation*}
This spectrum assumes the turbulence is critically balanced \cite{Goldreich1995-fv} with strong turbulence occurring below the curve $\kprltilde^{\rm CB}$, as motivated in Appendix C of Ref.~\cite{Schekochihin2022-nn}.

The above can be thought of as a simplistic model of fully imbalanced turbulence in the $\sigma_{\rm c}\to 1$ limit of a collection of $\bm{z}^\pm$ waves with $|\bm{z}^-|\ll |\bm{z}^+|$ undergoing a strong cascade.
This allows us to assume the turbulence consists of a collection of $\bm{z}^+$ Alfv\'en waves with dispersion relation $\omega = -k_\|\vA$.
Assuming a background magnetic field $\bm{B}_0=B_0\hat{\bm{z}}$, the velocity and magnetic fields of the waves are Alfv\'enically polarized ($\bm{e}_k\equiv \hat{\bm{k}}_\perp\times\hat{\bm{z}}$ with $\hat{\bm{k}}_\perp = \cos\phi\ \bm{e}_x+\sin\phi\ \bm{e}_y$), and thus using Ohm's law their electric field Fourier components are polarized in $\hat{\bm{k}}_\perp$: $\bm{E}(\bm{k},t)=\mathcal{A}_{\bm{E}}(\bm{k})e^{-i k_\|\vA t}\hat{\bm{k}}_\perp$.

From Ref.~\cite{Kennel1966-rx} it is shown that, in the limit of zero wave dampening, the equilibrium velocity distribution function $f_0$ of a collection of ions undergoes diffusion in velocity space due to ion-wave resonances of the form
\begin{eqnarray}
    \frac{\partial f_0}{\partial t} = \frac{1}{v_\perp}\frac{\partial}{\partial v_\perp}\left[v_\perp\left(D_{\perp\perp}\frac{\partial f_0}{\partial v_\perp}+D_{\|\perp}\frac{\partial f_0}{\partial v_\|}\right)\right]\nonumber\\
    +\frac{\partial}{\partial v_\|}\left(D_{\perp\|}\frac{\partial f_0}{\partial v_\perp}+D_{\|\|}\frac{\partial f_0}{\partial v_\|}\right)
\end{eqnarray}
with perpendicular diffusion coefficient (using the dispersion relation $\omega = -k_\|\vA$)
\begin{eqnarray}
    D_{\perp\perp} = \lim_{V\to\infty}\frac{\pi q^2}{m^2}\left(1+\frac{v_\|}{\vA}\right)^2\times\nonumber\\
    \sum_{n=-\infty}^{\infty}\int\frac{d\bm{k}}{V}\delta\left[k_\|(\vA+v_\|)+n\Omega_{i}\right]|\psi_{n,\bm{k}}|^2,
\end{eqnarray}
$D_{\perp\|}=D_{\|\perp}=-v_\perp D_{\perp\perp}/(\vA+v_\|)$ and $D_{\|\|}= (v_\perp/(\vA+v_\|))^2 D_{\perp\perp}$.
Here, $\psi_{n,\bm{k}} = E^+_{\bm{k}}J_{n-1}(z)+E^-_{\bm{k}}J_{n+1}(z)$ is the wave amplitude, $z=k_\perp v_\perp/\Omega_i$ is the argument of the Bessel functions and $E^\pm_{\bm{k}}\equiv (E_{\bm{k}x}\pm iE_{\bm{k}y})e^{\mp i\phi}/2 = \mathcal{A}_{\bm{E}}(\bm{k})/2$ after inserting the polarization, which simplifies the wave amplitude to $\psi_{n,\bm{k}} = \mathcal{A}_{\bm{E}}(\bm{k}) n J_{n}(z)/z$ after using the identity $J_{n-1}+J_{n+1}=2nJ_{n}/z$.
The sum over $n$ then collapses to the $n=\pm 1$ terms, as for $z\ll 1$ we have $n^2J^2_{n}/z^2 \approx 0$ for $|n|\neq 1$ and $J^2_{\pm 1}/z^2 \approx 1/4$.
With $\mathcal{A}_{\bm{E}}(\bm{k}) = (B_0/c)E_{\rm 2D}(k_\perp,k_\|)$, we can integrate over the spectrum.
For $n=\pm 1$, the $\delta$-function gives a resonance at parallel scales $|k^{(1)}_\||\equiv 1/(1+v_\|/\vA)$ with only ions with $v_\| > 0$ able to resonate; for low-$\beta$ ions to resonate with modes in the strong turbulence regime requires $(\ion{\rho}/\lambda)(\deltaup/\vthi)\geq 1/(1+v_\|/\vA)\approx 1$.
Performing the integral in this regime gives
\begin{eqnarray}\label{eq:diffusioncoeff}
    \frac{D_{\perp\perp}}{\ion{\Omega}\vthi}\approx\frac{\pi}{32}\left(1+\frac{v_\|}{\vA}\right)^2\left(\frac{\ion{\rho}}{\lambda}\right)\left(\frac{\deltaulambda}{\vthi}\right)^3\times\nonumber\\
    \left[5\left(1+\frac{v_\|}{\vA}\right)^2-3\left(\frac{\ion{\rho}}{\lambda}\right)^{-2}\left(\frac{\deltaulambda}{\vthi}\right)^{-2}\right]\propto \ion{\xi}^3F
\end{eqnarray}
where $\ion{\xi}\equiv (\ion{\rho}/\lambda)^{1/3}(\deltaulambda/\vthi)$ and $F = 5-3(\ion{\rho}/\lambda)^{-2}(\deltaulambda/\vthi)^{-2}$.
A similar calculation for smaller velocities such that $(\ion{\rho}/\lambda)(\deltaup/\vthi)< 1/(1+v_\|/\vA)$ gives $D_{\perp\perp}\propto \ion{\xi}^6(1+v_\|/\vA)^7$.

\begin{figure}
\centering
\includegraphics[scale=0.7]{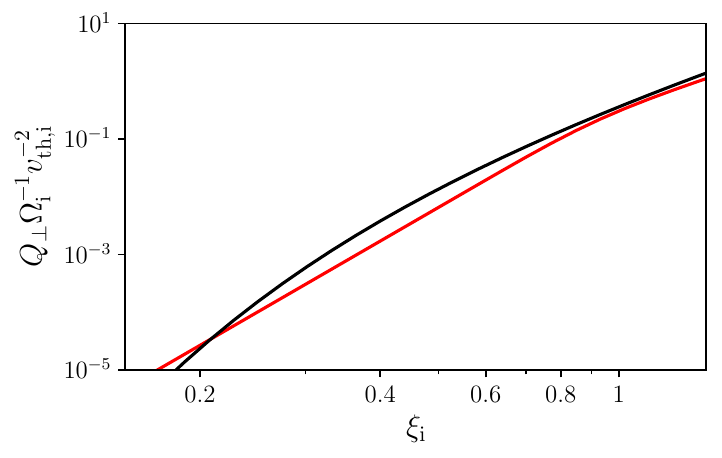}
\caption{\label{fig:heatingrate}
$Q_\perp$ calculated from the $v_\|$-averaged diffusion coefficient $\tilde{D}_{\perp\perp}$ (red) assuming a Maxwellian distribution with $\beta=0.1$ compared
to the standard SH formula (Eq. 1 of Letter with $c_1=1.0,c_2=1.2$, black).
This heating rate is similar to the exponential suppression of the SH model for $\xi_{\rm i}\lesssim 1$, and decreases slower for very small $\xi_{\rm i}\lesssim 0.2$ due to the $\ion{\xi}^6$ dependence of the diffusion coefficient.}
\end{figure}

We calculate the perpendicular heating rate numerically via $-2\pi\int_0^\infty dv_\perp \ v^2_\perp \tilde{D}_{\perp\perp}(\partial f(v_\perp)/\partial v_\perp)$ \cite{Arzamasskiy2019-qv,Cerri2021-xo}, where the distribution is assumed to be Maxwellian so that $\tilde{D}_{\perp\perp}\equiv \int_{-\infty}^\infty dv_\| D_{\perp\perp} e^{-v^2_\| / (\beta \vA^2)}$ is the Maxwellian-weighted diffusion coefficient and $f(v_\perp)=(\pi\vthi^2)^{-3/2}e^{-v^2_\perp / (\beta \vA^2)}$.
Figure \ref{fig:heatingrate} compares this calculated heating rate for a Maxwellian distribution with $\beta=0.1$ to the SH formula with exponential suppression (Eq.~\ref{eq:Qprp} with $c_1=1.0$ and $c_2=1.2$; these coefficients are larger than those seen in simulations and observations due to factors of order unity throughout our definitions of the spectrum and critical balance).
The suppression factor of the calculated heating rate is of a similar form to the exponential factor assumed by Ref.~\cite{Chandran2010-ow} for $\ion{\xi}\lesssim1$, but does not drop off as fast for very small $\ion{\xi}$ due to the $\ion{\xi}^6$ dependence of $D_{\perp\perp}$.
Despite this, the assumption of exponential suppression in Equation 1 of the Letter works well in describing the measured heating rates of ions.

Despite appearing to depend on $\ion{\rho}$ through the factor $\ion{\rho}/\lambda$, this expression is in fact ignorant of $\ion{\rho}$ physics, which appears only due to the normalization used as $\ion{\Omega}\vthi^2(\ion{\rho}/\lambda)(\deltaulambda/\vthi)^3 = \deltaulambda^3/\lambda\sim (z^+/z^-)\varepsilon$, the turbulent cascade rate that is constant throughout the inertial range.
For the same reason, the results in Fig.~\ref{fig:SH_heatingrate} do not depend significantly on the choice of $\lambda$, so long as it provides a reasonable estimate of the break scale caused by the helicity barrier.
Although we used the scale where $k_\perp\tilde{z}^+_{k_\perp}$ peaks for $\lambda$, we note that the empirical form for the helicity barrier break scale $\ion{\rho}/\lambda\propto (1-\sigma_{\rm c})^{1/4}$ \cite{Meyrand2021-ix,Squire2023-jn} also works well.
We note that performing the same calculation taking the assumptions of dynamic alignment \cite{Boldyrev2006-xd} into account gives the same scaling for $\ion{\xi}$, as we now have to take into account that it is $\deltaulambda^3 \sin\theta_\lambda / \lambda$ that is constant rather than $\deltaulambda^3/\lambda$ (where $\theta_\lambda$ is the angle between the two Els\"asser fields).

\section*{Supplemental Material}

\subsection*{The FLR-MHD Model}

We use the finite Larmor radius MHD (FLR-MHD) model \cite{Passot2018-yt,Schekochihin2019-al,Meyrand2021-ix}, a reduced fluid model that includes effects at scales both larger and smaller than $\ion{\rho}$ without needing to solve kinetic equations.
This theory can be derived from the Vlasov-Maxwell equations by assuming all quantities vary on timescales slower than the ion-cyclotron frequency, a strong background magnetic field, and that perturbations are elongated along magnetic field lines \cite{Howes2006-tq,Schekochihin2009-qo}.
Assuming a background magnetic field $\bm{B}_0=B_0\hat{\bm{z}}$ and taking the low-$\beta$ limit \cite{Zocco2011-zp,Schekochihin2019-al} gives the equations of FLR-MHD:
\begin{widetext}
\begin{align}
    \left(\frac{\partial}{\partial t}+\bm{u}_\perp\bm{\cdot}\bm{\nabla}_\perp\right)\frac{\delta n_e}{n_{0e}} &= -\frac{c}{4\pi e n_{0e}}\left(\frac{\partial}{\partial z}+\bm{b}_\perp\bm{\cdot}\bm{\nabla}_\perp\right)\bm{\nabla}_\perp^2A_\|+ \mathcal{D}_{6\nu}\frac{\delta n_e}{n_{0e}},\\
    \left(\frac{\partial}{\partial t}+\bm{u}_\perp\bm{\cdot}\bm{\nabla}_\perp\right)A_\| &= -c\frac{\partial \varphi}{\partial z}+\frac{cT_{0e}}{e}\left(\frac{\partial}{\partial z}+\bm{b}_\perp\bm{\cdot}\bm{\nabla}_\perp\right)\frac{\delta n_e}{n_{0e}} + \mathcal{D}_{6\eta}A_\|,\\
    \frac{\delta n_e}{n_{0e}}&=-\frac{Z}{\tau}\left(1-\hat{\Gamma}_0\right)\frac{e\varphi}{T_{0e}},
\end{align}
\end{widetext}
where $\tau=T_{0\text{i}}/T_{0\text{e}}$ is the ratio of the ion and electron temperatures, $\delta n_e/n_{0e}=\delta n_{\rm i}/n_{0\rm i}$ is the perturbed electron density (equal to the ion density by quasi-neutrality), $A_\|$ is the $\hat{\bm{z}}$ component of the magnetic vector potential, $\varphi$ is the electrostatic potential, $\bm{u}_\perp = (c/B_0)\hat{\bm{z}}\times\bm{\nabla}_\perp\varphi$ is the perpendicular $\ExB$ flow, and $\bm{b}_\perp = -\hat{\bm{z}}\times\bm{\nabla}_\perp A_\|/B_0$ is the magnetic field perturbation.
The gyrokinetic Poisson operator
$1-\hat{\Gamma}_0 = 1-I_0(\alpha)e^{-\alpha}$, where $I_0$ is the modified Bessel function of the first kind and $\alpha = -\ion{\rho}^2\bm{\nabla}^2_\perp/2$, captures the transition from an Alfv\'enic to a kinetic-Alfv\'en-wave cascade that occurs at $k_\perp\ion{\rho}\sim 1$.
In our simulations a hyperdiffusion operator $\mathcal{D}_{6\nu} = \nu_{6\perp}\bm{\nabla}_\perp^6 + \nu_{6z}\bm{\nabla}_z^6$ (and similarly for $\mathcal{D}_{6\eta}$) is used to dissipate energy above grid scales, but does not model any specific physical process; we set $\eta_{\perp,z} = \nu_{\perp,z}$ in all cases except balanced FLR-MHD, discussed below.

This model describes the evolution of the Alfv\'enic component of turbulence both above and below $\ion{\rho}$; the assumption of small $\beta$ leads to minimal coupling between this and the compressive component of the turbulence \cite{Schekochihin2009-qo}.
The FLR-MHD model is valid as long as the electron inertial length $d_{\text e} \ll \ion{\rho}$.
This is because $d_{\rm e}$ sets the length scale below which the magnetic field is no longer frozen into the electron flow; in this regime the equations are coupled to the electron distribution function and are no longer closed.
Further details on the properties of FLR-MHD are given in Ref.~\cite{Meyrand2021-ix}.

\subsubsection*{Eigenmodes and Conserved Invarients}
For a given wavenumber $\bm{k}$, the FLR-MHD model supports eigenmodes that travel parallel and anti-parallel to $\bm{B}_0$ given by
\begin{equation}
    \Theta^{\pm}_{\bm{k}} = -\Omega_{\rm i}\frac{v_{\rm ph}(k_\perp)}{k^2_{\perp}}\frac{\delta n_e}{n_{0e}}\mp\frac{A_\|}{\sqrt{4\pi m_{\rm i}n_{0\rm i}}},
\end{equation}
with frequency $\omega = \pm k_z v_{\rm ph}(k_\perp)\vA$.
The normalized phase speed $v_{\rm ph}(k_\perp)=k_\perp\ion{\rho}[(1/[1-\hat{\Gamma}_0]+Z/\tau)/2]^{1/2}$ reduces to 1 at $k_\perp\ion{\rho}\ll 1$ scales and is proportional to $k_\perp\ion{\rho}$ at $k_\perp\ion{\rho}\gg 1$ scales, reflecting the transition from Alfv\'en to kinetic-Alfv\'en waves.
These eigenmodes allow the definition of generalized Els\"asser variables $\tilde{\bm{z}}^\pm = \hat{\bm{z}}\times\bm{\nabla}_\perp\Theta^{\pm}$ that reduce to the standard Els\"asser variables of MHD ($\bm{z}^\pm = \bm{u}_\perp\pm \bm{B}_\perp/\sqrt{4\pi m_{\rm i}n_{0\rm i}}$) in the limit $k_\perp\ion{\rho}\ll 1$.

This model also has two conserved invariants:
the free energy
$E = \frac{1}{4}\sum_{\bm{k}}(|k_\perp\Theta^+|^2+|k_\perp\Theta^-|^2)$,
and the generalized helicity $\mathcal{H}  = \frac{1}{4}\sum_{\bm{k}}(|k_\perp\Theta^+|^2-|k_\perp\Theta^-|^2)/v_{\rm ph}(k_\perp)$. The helicity barrier results from the inability of the system to support a forward cascade of both $E$ and $\mathcal{H}$ simultaneously at $k_\perp\ion{\rho}\gg 1$, due to the scaling $v_{\rm ph}\propto k_\perp$ in this range \cite{Meyrand2021-ix}.

\subsection*{Additional Simulation Details}

We use \textsc{asterix} \cite{Meyrand2019-ms,Meyrand2021-ix,David2024-bm}, a modified verison of the pseudospectral code \textsc{turbo} \cite{Teaca2009-go}, to solve the equations of RMHD and FLR-MHD, which are advanced in time with a third-order modified Williamson algorithm (a four-step, low-order Runge-Kutta method \cite{Williamson1980-hz}). 
We force the imbalanced FLR-MHD and RMHD simulations with energy injection $\varepsilon$ and $\sigma_\varepsilon=0.88$, where $\sigma_\varepsilon = |\varepsilon_{\mathcal{H}}|/\varepsilon$ is the ratio of the helicity and energy injection rates.
In contrast, the balanced FLR-MHD simulation has no helicity injected into the system ($\varepsilon_\mathcal{H} = 0$).

\begin{figure}
\centering
\includegraphics[scale=0.7]{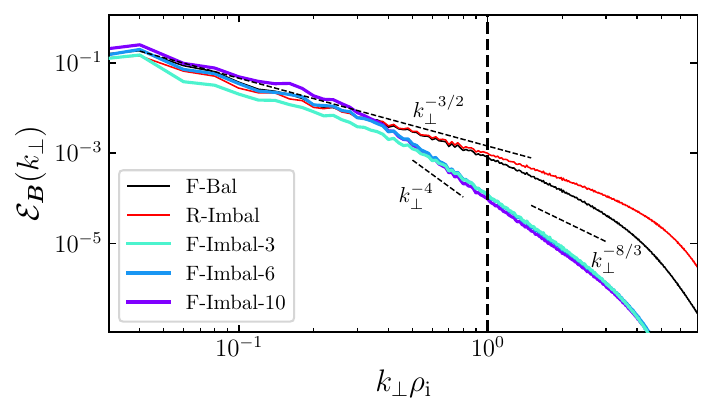}
\caption{\label{fig:mag-spec} Comparison of magnetic-field spectra from the different turbulence simulations in this Letter.}
\end{figure}

Figure \ref{fig:mag-spec} shows the magnetic-field spectra of the high-resolution simulations after refinement (discussed below), corresponding to the electric-field spectra in Figure 1 of the Letter.
All simulations exhibit an approximate $-3/2$ scaling for $k_\perp\ion{\rho}\lesssim 1$.
With the addition of a helicity barrier, the magnetic-field spectra show similar scalings to electric-field spectra, steeping to a $-4$ scaling around  $k_\perp\ion{\rho}\sim 1$.

\subsubsection*{Choice of Parallel Viscosity}

We use different choices of the parallel viscosity $\nu_{6z}$ in the different turbulent regimes to reflect how they arrive at a steady state.
For imbalanced RMHD turbulence, the majority of the energy cascade is able to reach small perpendicular scales and dissipate there.
However, due to critical balance this energy also reaches small parallel scales.
To ensure correct dissipation at small scales, we use a non-zero parallel viscosity.
For the helicity barrier, however, we are interested in how it affects the heating of test particles as it grows in time during the ``pseudo-stationary" phase, before it saturates unphysically on the parallel hyperdissipation.
We thus set $\nu_{6z}=0$ for the helicity barrier simulation, studying it well before it would saturate with $\nu_{6z}\neq 0$. 
Various tests (not shown) found that this produces results similar to using a non-zero $\nu_{6z}$.

\subsubsection*{Simulation Refinement}

A highly imbalanced forcing causes the turbulence to take longer to reach a steady-state, as the energy cascade is less efficient for the stronger Els\"asser field \cite{Lithwick2007-ao,Schekochihin2022-nn}.
The addition of a helicity barrier complicates this further as the system evolves through a ``pseudo-stationary" state where the balanced portion of the cascade dissipates at small perpendicular scales, whereas the imbalanced portion grows at scales $k_\perp\ion{\rho}\lesssim 1$ before saturating unphysically on parallel hyperdissipation \cite{Meyrand2021-ix,Squire2022-dm}.
To run a high-resolution simulation, such as the ones used in this work, to this saturation point would be time-consuming and computationally expensive.

Instead, we start from simulations with a resolution of $N_\perp^2\times N_z = 256^2\times 512$ and let these evolve to a steady-state (or pseudo-stationary state for simulations with a helicity barrier).
Once this point is reached, we refine the simulation by doubling the perpendicular resolution and decreasing the perpendicular viscosity $\nu_{6\perp}$, allowing the energy cascade to reach smaller scales before dissipating.
Once the cascade has reached a new steady-state, we repeat this refinement procedure until the desired resolution is reached, which is then used as the starting point for our test particle simulations.
The final resolutions are $N_\perp=1024,\ N_z = 512$ for the imbalanced RMHD and FLR-MHD cases, and $N_\perp=N_z=1024$ for the balanced FLR-MHD case (the reason for this is discussed below).

\begin{figure}
\centering
\includegraphics[scale=0.7]{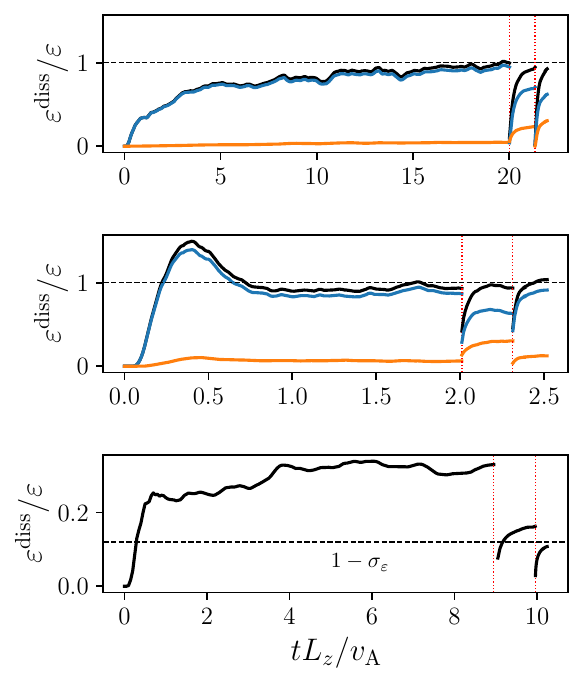}
\caption{\label{fig:refining} Time evolution of turbulent dissipation for the imbalanced RMHD (top), and the balanced and imbalanced FLR-MHD cases (middle, bottom respectively). 
Contributions to the parallel (orange) and perpendicular (blue) dissipation and their total (black) are normalized to the energy input $\varepsilon$. The red dotted lines show when the simulations were refined, doubling the perpendicular resolution from $256^2\times 512$ up to $1024^2\times 512$ for the imbalanced cases or $1024^3$ for the balanced case. Note the difference in scale on both axes; see text for discussion.}
\end{figure} 

\begin{figure*}
\centering
\includegraphics[scale=0.63]{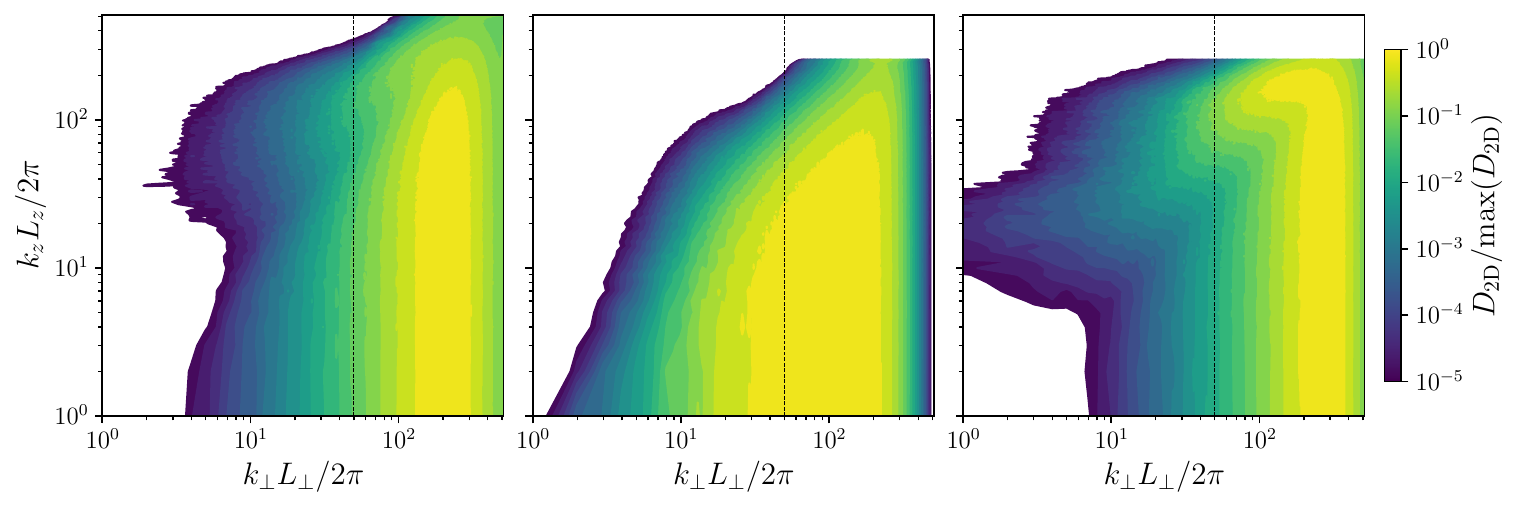}
\caption{\label{fig:diss-spec} Two-dimensional dissipation spectra $D_{2\rm D}(k_\perp,k_z)$ from the $1024^3$ balanced FLR-MHD simulation (left) and the $1024^2\times 512$ imbalanced FLR-MHD (middle) and RMHD (right) simulations, normalized to their maximum values.
The vertical dotted lines represent the test particle gyroradius scale $k_\perp\ion{\rho}=1$.}
\end{figure*}

Figure \ref{fig:refining} compares the time-evolution of dissipation in all balanced and imbalanced RMHD and FLR-MHD cases during the refinement process.
The imbalanced cases take longer to saturate compared to the balanced case for the reason discussed above.
As we increase the perpendicular resolution while refining, the cascade can reach smaller corresponding parallel scales due to critical balance, leading to an increase in parallel dissipation.
This effect is clearly seen in the balanced FLR-MHD case: the initial refinement in $N_\perp$ leads to an increase in $\varepsilon_z$, and the subsequent refinement in both $N_\perp$ and $N_z$ resulting in a decrease (due to dissipation scales occuring at larger $k_z$).
In the imbalanced RMHD and balanced FLR-MHD cases, all energy is dissipated near the grid at predominately small perpendicular scales with $\varepsilon^{\rm diss}\approx \varepsilon$.
In contrast, the helicity barrier in the imbalanced FLR-MHD case blocks the majority of the cascade, only letting the balanced portion of the flux ($\sim 1-\sigma_\varepsilon$) through to be dissipated at grid scales; the remainder is ``trapped" at large perpendicular scales.
The lower-resolution simulations allow more of the cascade to be dissipated than predicted as their dissipation scales are close to $\ion{\rho}$-scales; this effect decreases as the range between these scales increases with resolution, as seen by the lower $\varepsilon^{\rm diss}/\varepsilon$ at each refinement level, converging to $\varepsilon^{\rm diss}/\varepsilon=1-\sigma_{\varepsilon}$.

Figure \ref{fig:diss-spec} compares two-dimensional dissipation spectra $D_{2\rm D}(k_\perp, k_z)$ of all energy from the high-resolution balanced and imbalanced FLR-MHD simulation and the imbalanced RMHD simulation.
In the absence of a helicity barrier, energy is dissipated at grid scales in both the RMHD and FLR-MHD case.
In contrast, the presence of a helicity barrier causes a flat dissipation spectrum around $k_\perp\ion{\rho}\sim 1$, where the majority of the cascade is halted that causes the steep slope in the energy spectrum in this range.
Note that the dissipation spectra includes the balanced portion of the energy cascade let through by the helicity barrier, and thus shows dissipation at $k_\perp\ion{\rho}\gtrsim 1$ scales in the imbalanced FLR-MHD simulation; however, the majority of the imbalanced cascade is dissipated at $k_\perp\ion{\rho}\lesssim 1$ scales.

\subsubsection*{Viscosity in the Balanced FLR-MHD Case}

\begin{figure}
\centering
\includegraphics[scale=0.65]{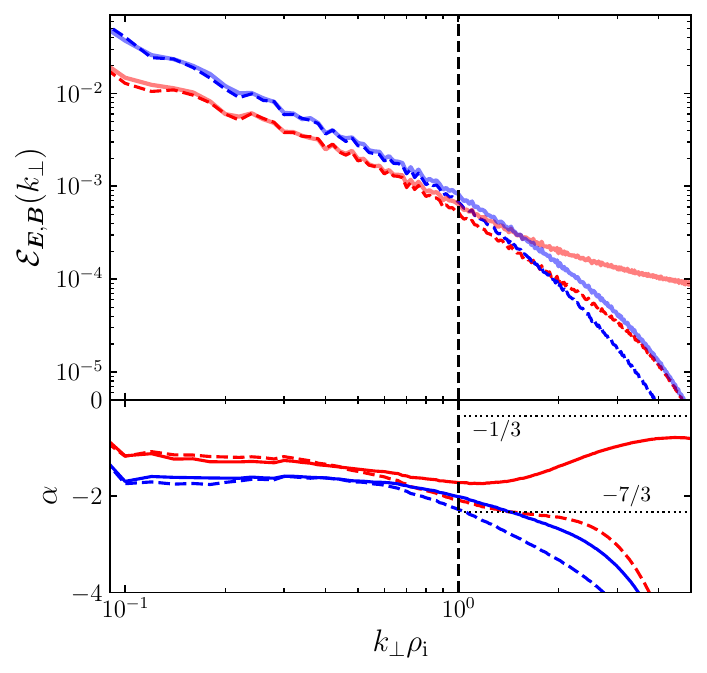}
\caption{\label{fig:flrbal-spec} Magnetic- (blue) and electric-field (red) spectra and spectral slope $\alpha$ (measured by fitting over a range $\pm\log_{10}(k_\perp)\approx 0.1$ around each point), highlighting the dependence on dissipation by viscosity. Dashed lines show the spectra from a $N_\perp^2\times N_z=1024^2\times 512$ balanced FLR-MHD simulation with viscosity dissipation activated, and solid lines from the $1024^3$ balanced simulation with $\nu=0$ (used in the main Letter).}
\end{figure}

A notable exception to the methods described above was the balanced FLR-MHD turbulence case.
Due to the transition from an Alfv\'enic to a kinetic-Alfv\'en-wave cascade at scales $k_\perp\ion{\rho}\gtrsim 1$ we expect the slope of the magnetic-field spectrum to steepen from $-5/3$ to $-7/3$, and the electric-field specturm to flatten from $-5/3$ to $-1/3$ \cite{Adkins2024-xf}.
However, when refining the balanced FLR-MHD simulation to a resolution of $1024^2\times 512$ with the method above we found that both the magnetic- and electric-field spectra were steeper than expected (dashed blue and red lines in Fig.~\ref{fig:flrbal-spec}, respectively), with the electric field reaching a slope of around $-7/3$.
These spectra (solid lines in Fig.~\ref{fig:flrbal-spec}) approach their theoretical predictions after refining the simulation to a resolution of $1024^3$ and switching off only the viscosity ($\nu_{\perp,z} = 0$ and $\eta\neq 0$), leaving the electric field unaffected by dissipation (a method used in hybrid-kinetic simulations \cite{Arzamasskiy2019-qv,Cerri2021-xo,Squire2022-dm}).
This steepening appears to be an unexpected nonlinear effect previously unseen in the FLR-MHD model.
Previous work has shown that the reconnection of current sheets by sub-$\ion{\rho}$ effects can lead to a steepening of the energy spectrum similar to what is seen in the transition range \cite{Mallet2017-mg,Loureiro2017-xo,Zhou2023-vm}; however, these require the electron inertial length $d_{\rm e}$ which is ordered out of FLR-MHD.
Whether what we observe here is related to this physics remains unclear, and more work is required to clarify the details of this mechanism.
Because of this, we integrate particles using the $\nu=0$ balanced FLR-MHD simulation in the main Letter to remove any influence on heating due to this effect. 

\begin{figure*}
\centering
\includegraphics[scale=0.72]{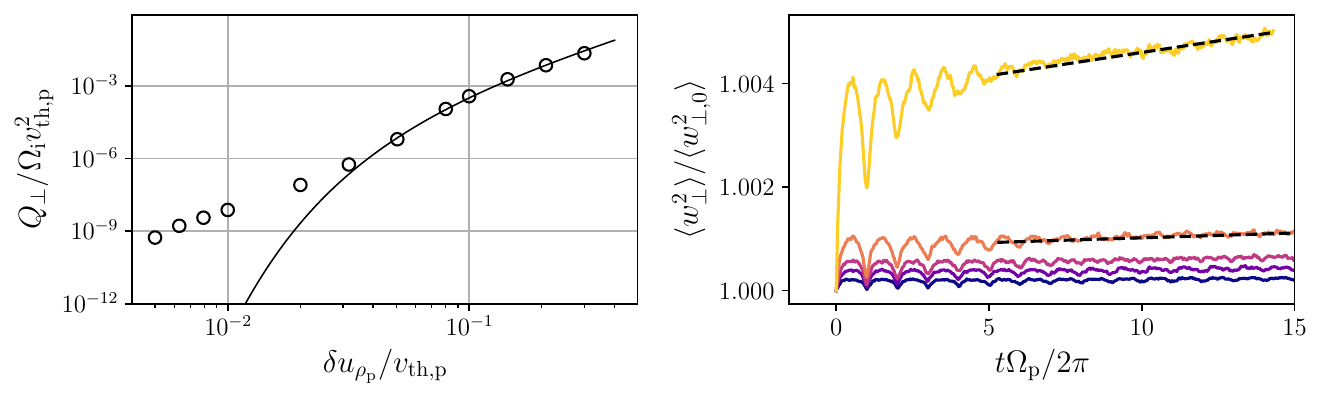}
\caption{\label{fig:heating-noise} Measured heating rate $Q_\perp$ of protons in $1024^3$ balanced FLR-MHD turbulence (left), demonstrating the greater-than-expected heating at low turbulent amplitudes. The solid line shows the best fit to the empirical $Q_\perp$ (Eq. 1 in Letter) for \flrbal \ with $c_1, c_2=1.83,0.18$.
These heating rates are measured using a linear fit (dashed lines; right) to the ensemble-averaged perpendicular energy per unit mass $\langle w^2_\perp\rangle$ for the protons with $5\times 10^{-3} \leq \explicitxiproton \leq 2\times 10^{-2}$ (dark to light).
The oscillations are clearly seen and have mostly decayed after approximately $5$ gyroperiods; $Q_\perp$ is measured from $5.25$ gyroperiods onward for these small-$\explicitxi$ particles.
For clarity, $\langle w^2_\perp\rangle$ is normalized to its initial values; the values used for the heating rate calculations are normalized to $\vA^2$.}
\end{figure*} 

\subsection*{Additional Test Particle Details}

In order to study particle heating, we introduced the ability to integrate test particles simultaneously with the fields solved by \textsc{asterix}. 
As in previous test particle simulations (e.g.~\cite{Lehe2009-up,Xia2013-ob}), the equations of motion of the particles are integrated using the Boris push \cite{Boris1970-sf} with electric and magnetic fields interpolated from the grid to their positions via the triangular-shaped-cloud (TSC) method \cite{Lehe2009-up,Birdsall2018-mv}.
The test particles are run simultaneously with the fluid code so that particles can use the current fields without needing to save the full history of the simulation.
The perpendicular fields are obtained via $\bm{B}_\perp = B_0\bm{b}_\perp$ and $\bm{E}_\perp = -\bm{u}_\perp\times\bm{B}/c$ (with $\bm{u}_\perp$ and $\bm{B}_\perp$ as defined above via $\varphi$ and $A_\|$ in FLR-MHD); any artificial parallel electric field component due to interpolation is then removed using the method of Ref.~\cite{Lehe2009-up}.
We calculate the required timestep via $\Delta t_{\rm particle} = \min\left(2\pi/(N_1\ion{\Omega}), \Delta t_{\rm fluid}/N_2\right)$ \cite{Lehe2009-up}, where $\Delta t_{\rm fluid}$ is the fluid timestep calculated by \textsc{asterix}.
Setting $N_1=50$ and $N_2=5$ was found to give accurate results; these results converge so long as $N_1\gtrsim 10$ (also seen by Ref.~\cite{Lehe2009-up}).

There are some important differences between RMHD and FLR-MHD that need to be taken into account when implementing test particles.
Firstly, the initial $\ion{\rho}/L_\perp$ of the particle distribution is no longer a free parameter due to the inclusion of kinetic scales in the FLR-MHD model (in comparison to RMHD, which assumes $\ion{\rho}\to 0$).
We fix this parameter during the initialization of particles in an FLR-MHD simulation.
Secondly, the inclusion of $k_\perp\ion{\rho}\gg 1$ scales introduces a electric field component parallel to $\bm{B}=\bm{B}_0+B_0\bm{b}_\perp$ that arises from the conservation of magnetic flux in gyrokinetics \cite{Schekochihin2009-qo,Cerri2021-xo}, given by
\begin{equation*}
    E_\| = \bm{E}\cdot\hat{\bm{b}} = -\left(\frac{\partial}{\partial z}+\bm{b}_\perp\cdot\nabla\right)\left(\frac{T_{0e}}{e}\frac{\delta n_e}{n_{0e}}\right)
\end{equation*}
where $\hat{\bm{b}}$ is the unit vector along $\bm{B}$.
We include this term when interpolating fields to the particles, although it is generally small in comparison to $\bm{E}_\perp$.

\subsubsection*{Testing of Implementation}

After its implementation, the particle code was extensively tested to verify its correctness.
Simple cases with known analytical properties (such as motion in constant $\bm{E}_\perp$ and $\bm{B}_0$ fields as well as conservation of energy in $\bm{E}=-\nabla\varphi$ fields) are reproduced with a relative error of less than $10^{-3}$.
The implementation of the TSC interpolation scheme was found to exhibit second-order convergence up to a resolution of $128^3$, before additional errors dominate (e.g., from timestepping).
The accuracy of this scheme was tested by comparing the use of grid-interpolated fields to those at the particle position via more complicated analytical fields (such as a linear combination of sinusoidal electromagnetic modes); for a grid resolution of $128^3$ or greater, the maximum relative error in energy evolution is less than $10^{-2}$.

The convergence of statistics with particle number was tested by running cohorts of $N$ particles (with $N$ ranging from 100 to $10^6$ in powers of 10) in the same $128^3$ balanced RMHD turbulence simulation.
The relative error in the statistics of an $N=10^5$ and $N=10^6$ run is $\lesssim 1\%$; we use $N=10^6$ particles to allow for better statistics when calculating the velocity distribution functions.

\begin{figure*}
\includegraphics[scale=0.75]{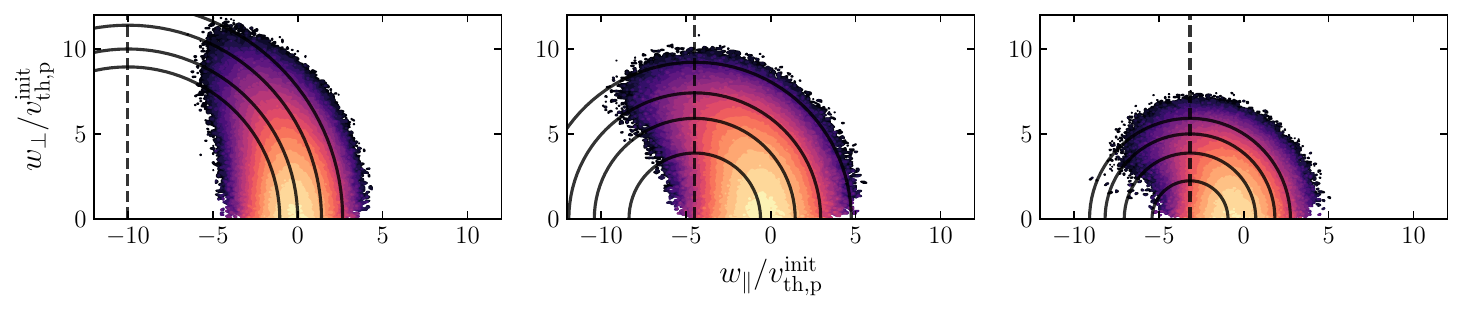}
\caption{\label{fig:beta-dep-heating} Comparison of two-dimensional velocity distributions $f(w_\perp, w_\|)$ of protons with $\proton{\hat{\xi}}\approx 0.7$ from \flrimbalsixlowbeta~(left), \flrimbalsix~(middle), and \flrimbalsixhighbeta~(right; same as Fig. 4 of main paper). Solid contours represent constant-energy contours (Eq.~\ref{eq:constantenergy}) in the frame of the wave, centred on the phase speed $\vph=-\vA$ (dashed line).}
\end{figure*}

\subsubsection*{Particle Heating at Small Turbulent Amplitudes}

Ions experience a greater-than-expected heating rate than the empirical exponential suppression in $Q_\perp = \ion{\Omega}\vthi^2\hat{c}_1\ion{\hat{\xi}}^3 e^{-\hat{c}_2/\ion{\hat{\xi}}} $ (Eq. 1 of Letter) when initialized with very low $\explicitxi$.
In Figure \ref{fig:heating-noise} we compare the measured heating rates of protons with $5\times 10^{-3} \leq \explicitxiproton \leq 1\times 10^{-2}$ interacting with the same $1024^3$ balanced FLR-MHD turbulence simulation as those with $0.02 \leq \explicitxiproton \leq 0.3$ (\flrbal \ from the main Letter).
Due to the low-amplitude turbulence at $\proton{\rho}$-scales, the evolution of $\langle w^2_\perp \rangle$ for these protons exhibits oscillations from the large-scale $\ExB$ flow that introduces 
some ambiguity in measuring the heating rate; these oscillations are reduced with increasing $\explicitxiproton$ as they experience greater heating (e.g., Figure 2 of the main Letter).
We find that the oscillations decay after $\sim 5$ gyroperiods for ions initialized with $\deltaui/\vthi < 0.02$, $\sim 3$ for those with $0.02 \leq \deltaui/\vthi \leq 0.05$, and $\sim 0.5$ for those with $\deltaui/\vthi > 0.05$.
To remove any contributions from the oscillations when measuring the linear growth of $\langle w^2_\perp \rangle$, we choose the start time of the linear fit for $Q_\perp$ to be $5.25$, $3.25$, and $0.75$ gyroperiods respectively for the above cases (with the quarter-gyroperiod offset ensuring we are in the middle of any further oscillation cycle).
Despite this, the measured heating rate is greater than that expected by exponential suppression ($Q_\perp/\proton{\Omega}\vthp^2 \lesssim 10^{-12}$ for $\explicitxiproton \lesssim 10^{-2}$).
This effect may be physical or numerical; more work is needed to understand its origins.

\subsubsection*{Beta-dependent Perpendicular Heating in Imbalanced FLR-MHD Turbulence}

In Fig.~\ref{fig:beta-dep-heating}, we show how the diffusion of the proton velocity distribution function in imbalanced FLR-MHD turbulence depends on $\proton{\beta}=\vthp^2/\vA^2$.
Quasi-linear theory predicts that the evolution of the distribution function $f(w_\perp, w_\|)$ is a diffusion along constant-energy contours
\begin{equation}
    K' = \frac{\proton{m}}{2}\left[w^2_\perp + (w_\| -\vph)^2\right]\label{eq:constantenergy}
\end{equation}
within the frame of the dominant waves that the ions interact with \cite{Kennel1966-rx,Stix1992-bo}.
In Fig.~\ref{fig:beta-dep-heating}, these are circles centred on the phase speed of Alfv\'enic fluctuations parallel to the magnetic field.
At smaller $\proton{\beta}$, the contours of $K'$ are nearly vertical for ions with small $w_\perp$, with diffusion then leading to an increase in perpendicular energy and a greater measured heating rate (as seen in Figure 3 of the Letter).

\end{document}